\documentclass{article}

\usepackage{amsmath,graphicx,subfig}
\usepackage{amssymb,amsfonts}
\usepackage{overpic}
\usepackage[cal=boondox]{mathalpha}
\usepackage{bm}
\usepackage{enumitem}
\graphicspath{ {./images/} }
\usepackage{multirow}
\usepackage{array}
\usepackage{mdwmath}
\usepackage{mdwtab}
\usepackage{booktabs}
\usepackage[nodisplayskipstretch]{setspace} 
\usepackage[ruled,vlined,lined,commentsnumbered]{algorithm2e}
\usepackage[compact]{titlesec}

\usepackage{arxiv}

\usepackage[utf8]{inputenc} 
\usepackage[T1]{fontenc}    
\usepackage{hyperref}       
\usepackage{url}            
\usepackage{booktabs}       
\usepackage{amsfonts}       
\usepackage{nicefrac}       
\usepackage{microtype}      
\usepackage{lipsum}		
\usepackage{graphicx}
\usepackage{doi}
\usepackage{csquotes}

\usepackage{color,array}
\usepackage{appendix}

\usepackage{amsmath,amsfonts,amssymb}
\usepackage{algorithmic}
\usepackage{bm}
\usepackage{balance}
\usepackage{algorithmic}
\usepackage[ruled,vlined,lined,commentsnumbered]{algorithm2e}
\usepackage{mathtools}
\usepackage{enumitem}
\usepackage{multirow}

\usepackage{tikz}
\usepackage{textcomp}

\title{Emerging Approaches for THz Array Imaging: A Tutorial Review and Software Tool}

\author{\href{https://orcid.org/0000-0002-3388-4805}{\includegraphics[scale=0.06]{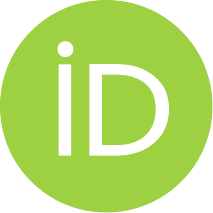}\hspace{1mm}Josiah W. Smith} \\
	Department of Electrical and Computer Engineering\\
	The University of Texas at Dallas\\
	Richardson, TX 75080 \\
	\texttt{josiah.smith@utdallas.edu} \\
	\And
	\href{https://orcid.org/0000-0001-7229-1765}{\includegraphics[scale=0.06]{orcid.pdf}\hspace{1mm}Murat Torlak}\thanks{This work was supported in part by Texas Instruments through the Foundational Technology Research Centre and the Texas Analog Center of Excellence.} \\
	Department of Electrical and Computer Engineering\\
	The University of Texas at Dallas\\
	Richardson, TX 75080 \\
	\texttt{torlak@utdallas.edu} 
}

\date{}

\renewcommand{\shorttitle}{Emerging Approaches for THz Array Imaging: A Tutorial Review and Software Tool}

\hypersetup{
pdftitle={Emerging Approaches for THz Array Imaging: A Tutorial Review and Software Tool},
pdfsubject={eess.SP, cs.AI},
pdfauthor={Josiah W.~Smith},
pdfkeywords={5G, 6G, computer vision, concealed item detection, deep learning, internet of things, synthetic aperture radar, terahertz imaging, submillimeter-wave imaging, submillimeter-wave technology, super-resolution, millimeter-wave imaging},
}

\begin{document}
\maketitle

\begin{abstract}
Accelerated by the increasing attention drawn by 5G, 6G, and Internet of Things applications, communication and sensing technologies have rapidly evolved from millimeter-wave (mmWave) to terahertz (THz) in recent years. 
Enabled by significant advancements in electromagnetic (EM) hardware, mmWave and THz frequency regimes spanning 30 GHz to 300 GHz and 300 GHz to 3000 GHz, respectively, can be employed for a host of applications. 
The main feature of THz systems is high-bandwidth transmission, enabling ultra-high-resolution imaging and high-throughput communications; however, challenges in both the hardware and algorithmic arenas remain for the ubiquitous adoption of THz technology. 
Spectra comprising mmWave and THz frequencies are well-suited for synthetic aperture radar (SAR) imaging at sub-millimeter resolutions for a wide spectrum of tasks like material characterization and nondestructive testing (NDT). 
This article provides a tutorial review of systems and algorithms for THz SAR in the near-field with an emphasis on emerging algorithms that combine signal processing and machine learning techniques. 
As part of this study, an overview of classical and data-driven THz SAR algorithms is provided, focusing on object detection for security applications and SAR image super-resolution. 
We also discuss relevant issues, challenges, and future research directions for emerging algorithms and THz SAR, including standardization of system and algorithm benchmarking, adoption of state-of-the-art deep learning techniques, signal processing-optimized machine learning, and hybrid data-driven signal processing algorithms. 
Further investigation into these directions will facilitate the transformation of THz SAR from theory to practical application and improve numerous sensing modalities. 
Finally, we introduce an interactive user interface and accompanying software toolbox to extend our review of THz SAR concepts, enable rapid system prototyping, accelerate algorithm development, and provide efficient dataset generation for emerging data needs. 
\end{abstract}


\section{Introduction}
\label{sec:intro}
The last decade has seen tremendous innovations in millimeter-wave (mmWave) and terahertz (THz) technologies for sensing and communication. 
mmWave wave frequencies, recognized here as 30 GHz to 300 GHz, offer a safe solution to many imaging and sensing problems due to their semi-penetrating, non-ionizing nature. 
With the perpetual demand for higher resolution sensing and increased throughput communications, THz frequencies, spanning from 300 GHz to 3000 GHz, are a suitable solution offering large bandwidths and enabling sub-millimeter resolution imaging. 
However, the evolution of system capabilities towards high-bandwidth mmWave and THz hardware has required innovation across several radio frequency (RF) domains. 

Imaging with THz waves was first explored in the 1990s \cite{hu1995imaging} and has been investigated for several decades \cite{mittleman2018twenty}.
In the early 2000s, the development of a fast electro-optical detector capable of working at video rate enabled the first two-dimensional (2-D) and three-dimensional (3-D) THz imaging systems \cite{jiang2000improvement}. 
Soon after, THz imaging was employed for applications including quality control \cite{jordens2008detection,wang2017emerging}, medical diagnosis \cite{humphreys2004medical,fitzgerald2006terahertz}, astronomy \cite{siegel2002terahertz,emerson1997work}, and threat detection \cite{davies2008terahertz,shen2005detection,kawase2003non,liu2007terahertz}.
Over the past decade, the application of THz systems has been extended to pharmaceutics \cite{shen2011terahertz}, art conservation \cite{fukunaga2016thz}, and automotive imaging \cite{daniel2018application,vizard2015low}.
However, innate hardware complexity hinders the progress of THz systems, and greater attention has been paid to mmWave technology. 

mmWave devices, the lower-frequency counterpart to THz, have emerged as a common choice for many applications due to their increasing availability and relatively high capacity, compared with the microwave/radio region \cite{soumekh1998wide}. 
Borrowing from microwave radar, mmWave systems have demonstrated tremendous success in synthetic aperture radar (SAR) imaging \cite{yanik2020development,sheen2001three}, a remote sensing technique developed in the 1950s for defense applications as an alternative to optical systems that support any lighting conditions and are capable of detecting hidden objects \cite{batra2021short,soumekh1999synthetic}. 
However, many mmWave SAR systems operate at short distances relative to the synthetic aperture size, knows as the near-field.
This requires delicate handling of the received signal for high-fidelity image recovery due to the curvature in the wavefront \cite{yanik2019sparse,smith2022efficient}. 
SAR imaging provides high cross-range resolution by synthesizing a large aperture using a smaller antenna aperture and processing subsequent measurements across time and space. 
For near-field SAR, the radar is mounted on a mobile platform close to the target and is scanned across a given space to create a synthetic aperture \cite{yanik2020development,smith2020nearfieldisar}. 
Backscattered reflections from the target contain high-fidelity spatial and temporal signatures of the illuminated target, which can be extracted through various means \cite{mohammadian2019sar,hangyo2005terahertz}. 
Near-field mmWave imaging systems have been employed in applications spanning concealed weapon detection \cite{liu2019concealed,yanik2019near}, automotive and UAV SAR \cite{iqbal2021realistic}, and human-computer interaction (HCI) \cite{smith2021An,kim2017staticgesture,smith2021sterile}. 
Spatial resolution is a crucial metric of any SAR system, where resolution is defined as the minimum resolvable distance between two scatterers in the scene using the half-power pulse width criterion \cite{soumekh1999synthetic}. 
The carrier frequency, system bandwidth, and aperture size directly affect the spatial resolution in the range and cross-range directions, respectively \cite{yanik2020development}. 
Whereas the cross-range direction is defined as parallel to the scanning direction, the range direction is typically defined as perpendicular to the direction of movement \cite{batra2021short}. 
Hence, as finer spatial resolution requires higher frequency and bandwidth, THz radar appears well suited as an improved alternative to mmWave for imaging applications; however, there are some trade-offs to consider.

In addition to the increased system complexity compared to mmWave systems, THz electromagnetic (EM) waves suffer from higher free space path loss \cite{rappaport2018wireless} and atmospheric absorption \cite{slocum2013atmospheric}, impairing the propagation distance and limiting some applications.
However, THz EM waves are preferable for material characterization because they enable excitation of vibrational mode in certain materials \cite{baxter2011terahertz}.
Therefore, THz spectral imaging can recover both the physical appearance of an object and its material composition. 
mmWaves, by comparison, suffer from poor spectral features for materials of interest in threat detection, such as explosives \cite{liu2007terahertz}. 
Whereas mmWave boasts a higher signal-to-noise ratio (SNR), the bandwidth of THz systems is approaching an excess of 400 GHz, which is larger than the entire mmWave spectrum \cite{batra2021Submm}.  
Both mmWave and THz systems typically employ transmission voltages on the order of millielectron volts, a significant reduction from harmful X-rays \cite{siegel2002terahertz}, and are considered highly safe in applications involving humans and other living organisms. 
Compared to infrared and visual regions, which are considered non-penetrating, both THz and mmWave waves offer suitable penetration for short-range applications, such as concealed item detection, object detection, and non-destructive testing (NDT)  \cite{batra2021short}. 

Algorithms for SAR imaging on both THz and mmWave systems have been developed for various common synthetic aperture geometries such as linear \cite{maisto2021sensor,smith2021An}, planar \cite{sheen2001three,yanik2020development,lopez20003,mohammadian2019sar,gao2018_1D_MIMO}, circular \cite{gao2016efficient,gezimati2022circular}, and cylindrical \cite{gezimati2023curved,gao2018cylindricalMIMO,smith2020nearfieldisar,detlefsen2005effective,berland2021cylindricalMIMOSAR}. 
Additionally, algorithms have been developed for SAR under irregular or pseudorandom geometries found in emerging technologies such as freehand imaging, UAV imaging, and automotive SAR \cite{smith2022efficient,alvarez2021freehand}. 
A typical approach for efficient SAR imaging solves the inverse scattering problem in the spatial wavenumber domain using a fast Fourier transform (FFT) approach and Stolt interpolation to account for the spherical curvature of the wavefront in the near-field, known as the range migration algorithm (RMA) \cite{yanik2019sparse,yanik2019cascaded,lopez20003,detlefsen2005effective}. 
However, variants such as the chirp scaling algorithm (CSA) \cite{wang2020AnEfficientCSA} and phase-shift migration (PSM) \cite{gao2019implementation,gao2020study} address efficiency and numerical accuracy, respectively. 
Additionally, compressed-sensing (CS) techniques have been applied to SAR, enabling high-resolution imaging with sub-Nyquist sampling rates by leveraging solution sparsity at the cost of increased computational burden \cite{molaei2021fourier,qiao2015compressive,helander2017compressive}. 
Recently, data-driven algorithms have emerged as a promising solution to SAR image super-resolution for THz \cite{fan2021fast,li2020adaptive} and mmWave \cite{huang2019through,smith2023dual_radar,cheng2020compressive,dai2021imaging,wang2021tpssiNet,smith2022ffh_vit} imaging, among other tasks, including hand gesture recognition \cite{kim2017staticgesture,smith2021sterile}, hand tracking \cite{smith2021An}, and threat detection \cite{yuan2018a,xiao2018r,xu2022YOLOMSFG,liu2019concealed,wei2021_3DRIED,lopez2017deep,lopez2018using,tapia2016detection}. 
Although deep learning algorithms quickly dominated the computer vision domain, their effectiveness on EM signals and images has only recently been explored. 
Therefore, as techniques continue to evolve, greater attention needs to be paid towards streamlining innovation, creating meaningful benchmarks, and identifying and quantifying underlying mechanisms that drive algorithm performance. 

The conventional backbone data-driven image processing is the convolution operation. 
Backpropagation-trained convolutional neural network (CNN) approaches gained momentum only a decade ago with the introduction of AlexNet \cite{krizhevsky2012imagenet}. 
Since then, a wide variety of CNN-based architectures have been introduced like VGGNet \cite{simonyan2014two_vggnet}, InceptionNet \cite{szegedy2015going_inception}, ResNet \cite{he2016deep_resnet}, ResNeXt \cite{xie2017aggregated_resnext}, DenseNet \cite{huang2017densely_densenet}, MobileNets \cite{howard2017mobilenet,sandler2018mobilenetv2}, EfficientNet \cite{tan2019efficientnet}, and RegNet \cite{radosavovic2020designing_regnet} emphasizing various architecture and implementation performance characteristics. 
By leveraging the sliding window convolution strategy, an operation intrinsically related to visual processing and boasting translational equivariance, CNN-based algorithms have produced many breakthroughs in image processing. 
However, the parallel field of natural language processing (NLP) took a different path from the CNN, starting with recurrent neural networks (RNN) and later Transformer techniques \cite{vaswani2017attention}. 
Recent research has introduced Vision Transformer (ViT) architectures \cite{dosovitskiy2020image_ViT,mehta2021mobilevit,liu2021swin} that leverage the self-attention mechanism found in NLP Transformers for vision tasks and have demonstrated competitive computational efficiency and accuracy. 
Both CNN \cite{fan2021fast,li2020adaptive} and ViT \cite{zheng2021dynamic,dong2021exploring,smith2022ffh_vit} approaches have been applied to THz and mmWave sensing tasks and will be detailed at length in this paper. 

Training reliable data-driven algorithms relies heavily on the quantity and quality of training data. 
This is a crucial issue for near-field SAR image processing as benchmark datasets are yet to be established, and current research relies on custom generation of datasets \cite{gao2018enhanced,fan2021fast,smith2021An,sharma2023super}. 
However, generating meaningful mmWave and THz data requires an in-depth understanding of the scattering phenomena and reconstruction algorithms.
There is a significant need for user-friendly data creation tools to enable the development of data-driven algorithms for SAR image processing. 
While this article aims to provide a gentle introduction to the near-field array imaging signal model and image reconstruction process for common synthetic aperture geometries, we also introduce a toolbox to increase the accessibility of SAR concepts and data generation. 
The proposed simulation tool allows full control over the virtual hardware setup, image recovery process, and dataset creation, and it is the first simulation tool with built-in capabilities for generating ground-truth data required for supervised learning. 
Earlier versions of the our open-source software platform have been employed in existing research for prototyping \cite{gezimati2022circular,gezimati2023curved,smith2020nearfieldisar}, algorithm development \cite{smith2022efficient}, and dataset generation \cite{smith2021An,smith2022ffh_vit,vasileiou2022efficient,smith2023dual_radar,smith2022novel}. 

The goals of this paper are as follows:
\begin{itemize}
    \item[1)] We review mmWave and THz radar signaling methods and modulation schemes for realizing high-bandwidth array imaging systems. 
    \item[2)] We introduce four common near-field SAR scanning regimes, discuss the underlying geometry, and derive efficient FFT-based image reconstruction algorithms. 
    In particular, we comprehensively detail algorithm implementation and discuss the signal intermediate steps of the reconstruction that can be leveraged for improving data-driven SAR image processing. 
    \item[3)] We discuss ongoing research efforts for mmWave and THz imaging circuits and systems. 
    As SAR applications continue to evolve beyond laboratory and controlled environments, understanding system limitations and design principles will becoming increasingly valuable for realizing novel systems in constrained applications. 
    \item[4)] We detail various approaches for deep learning-based SAR image processing at different stages in the image reconstruction process and discuss key insights that can be leveraged for future technological prospects. 
    Specifically, we discuss training regimes and the reliance on synthetic data; review challenges for designing neural network-based solutions such as complex-valued data representation and neural network design; and explore opportunities for future efforts including further investigation into weighted Frech\'et mean (wFM), standardization of radar benchmark datasets, development of end-to-end backpropagation-based learning algorithms, and exploration of neural network architectures beyond convolution.  
    \item[5)] We introduce an interactive user interface and API for near-field mmWave and THz SAR to improve the accessibility of SAR concepts and the availability of meaningful training data\footnote{Instructions for downloading and installing the proposed toolbox can be found in Appendix \ref{app:documentation}.}.
    \item[6)] Using the proposed software toolbox, we create a training dataset consisting of realistic SAR images with corresponding ground-truth labels and train a complex-valued neural network for the image super-resolution task. 
    The model is applied to real SAR images yielding a high-resolution reconstruction while retaining sharp edges and fine details, demonstrating the feasibility of applying networks trained on the synthetic data for real-world applications. 
\end{itemize}

The remainder of this paper is formatted as follows. 
In Section \ref{sec:sar_algorithms}, we introduce common signaling techniques and algorithms for near-field mmWave and THz SAR imaging. 
Imaging systems spanning GHz to THz frequencies are detailed in Section \ref{sec:sar_systems}. 
Section \ref{sec:ml_sar_algorithms} reviews the emerging applications of data-driven algorithms to SAR imaging, details the challenges of such approaches, and outlines future directions in this field. 
Finally, Section \ref{sec:thz_sim} introduces a novel interactive user interface for mmWave and THz imaging, prototyping, and dataset generation, followed by conclusions. 

\section{Signaling and Algorithms for Synthetic Aperture Radar Imaging}
\label{sec:sar_algorithms}

\begin{figure*}[ht]
    \centering
    \includegraphics[width=0.95\textwidth]{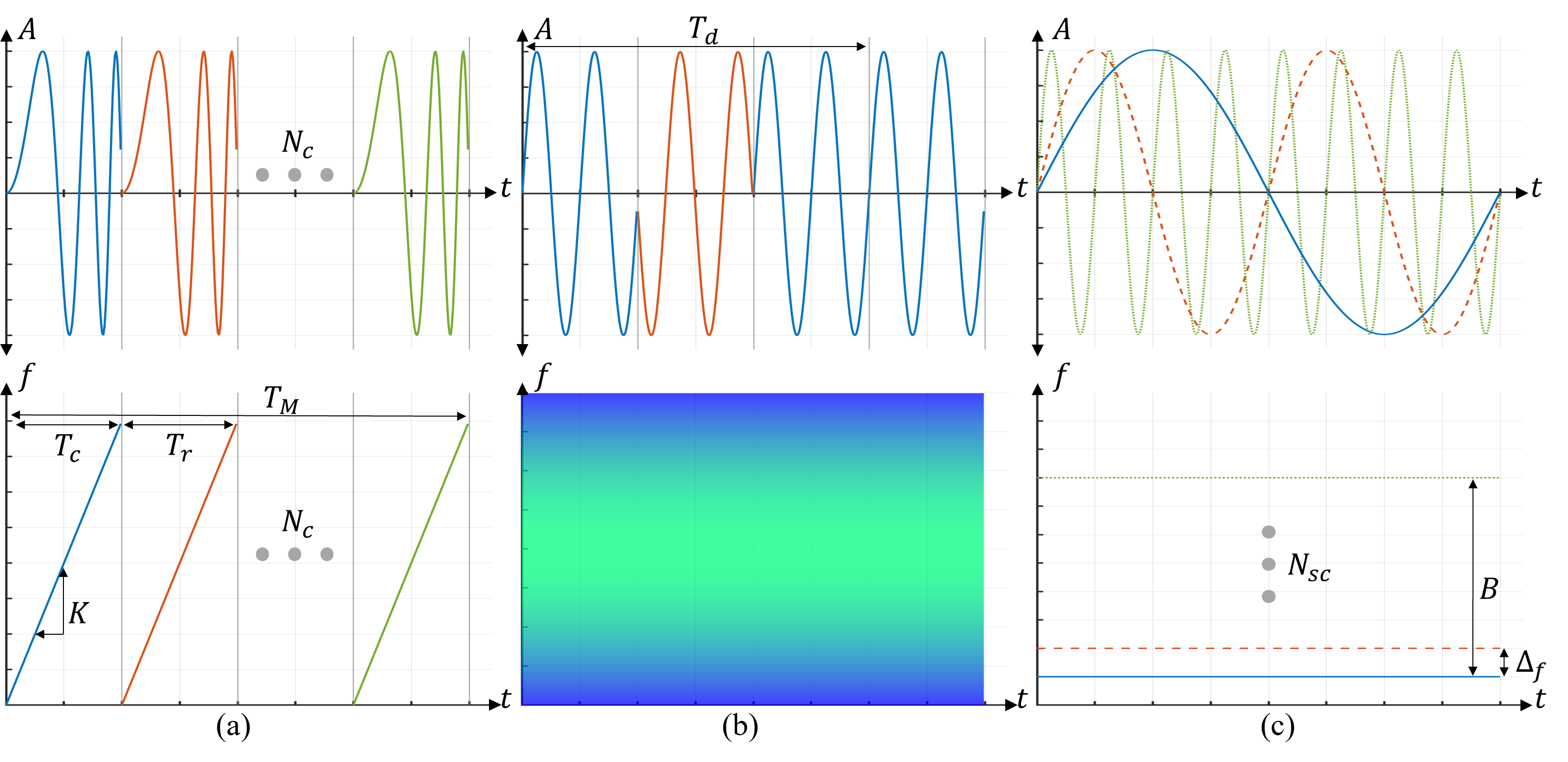}
    \caption{A summary of the common modulation schemes in the time domain (top) and corresponding frequency domain (bottom): (a) The FMCW chirp signal typically consists of a sequence of linearly sloped ramps (plotted in different colors). (b) The PMCW signaling scheme modulates the phase of a signal using binary orthogonal coding techniques (symbols are indicated with different colors), and the spectrum is shown with Hann windowing. (c) In OFDM radar, orthogonal subcarriers are transmitted to carry encoded information using the entire available spectrum. $A$: amplitude, $t$: time, $f$: frequency. Other symbols follow Section \ref{subsec:modulation_schemes}.}
    \label{fig:mod_schemes}
\end{figure*}

In this section, we detail signaling preliminaries, system models, and imaging algorithms for near-field SAR imaging. 
Common SAR systems use ultra-wideband (UWB) transceivers operating at bandwidths on the scale of several to tens of GHz. 
Such systems can be used to measure the reflection coefficient and spatial reflectivity function of a target across individual frequency points. 
Various signal modulation schemes have been explored throughout the literature in conjunction with different synthetic aperture geometries. 

\subsection{High-Bandwidth mmWave and THz Signal Modulation Schemes}
\label{subsec:modulation_schemes}

Frequency-modulated continuous-wave (FMCW) signaling is frequently employed in many commercially available radars, because of its low power consumption and ability to rely on low sample rate analog-to-digital converters (ADCs). 
In an FMCW transceiver, the transmitted signal, known as a chirp, consists of a single-tone sinusoid, whose frequency increases linearly with time. 
The reflected signal is received and mixed with the transmitted signal, which downconverts it to the baseband \cite{roos2019radar}. 
The resulting signal, known as the radar beat signal, contains a sinusoidal signal for each reflection of the target, allowing spatial information to be extracted by Fourier analysis. 
Since the beat signal is typically low-bandwidth, in the megahertz range, it can be sampled by a low-cost ADC. 
A typical operation involves the rapid transmission of a sequence of chirps, as shown in Fig. \ref{fig:mod_schemes}a, which effectively decouples the Doppler frequency from the range information \cite{winkler2007range}, allowing for multiple targets to be distinguished from a single measurement. 

FMCW signaling is characterized by the key parameters shown in Fig. \ref{fig:mod_schemes}a and detailed here. 
The range resolution $\delta_R$ is dictated by the total bandwidth $B$, which is a function of the chirp duration $T_c$ and ramp slope $K$, and the maximum resolvable range $R_\text{max}$ is limited by the fast-time sampling frequency $f_S$. 
The chirp repetition frequency, also known as the pulse repetition frequency $f_\text{PRF} = 1/T_r$, determines the unambiguous maximum Doppler velocity $v_\text{max}$ and the Doppler resolution $\delta_v$ is related to the total measurement time $T_M = N_c T_r$ \cite{roos2019radar,smith2021An}. 

A digital radar alternative to FMCW is phase-modulated continuous-wave (PMCW) signaling, which consists of a sequence of phase-modulated signals \cite{bourdoux2016PMCW}. 
Digital radar implementations rely on software-defined radio (SDR) architectures, which provide additional flexibility, adaptability, and robustness compared to single-carrier radar systems. 
A typical PMCW transmitter employs a binary-symbol approach to modulate the carrier frequency $f_c$ with $0^\circ$ or $180^\circ$ phase shifts using various binary sequences to achieve orthogonality.
Hence, an advantage of PMCW radar is that it can achieve transmit orthogonality through coding and waveform design \cite{roos2019radar}. 
The received signal is correlated with the transmitted signal for range processing.
However, Doppler shift adversely affects localization using PMCW radar and must be accounted for by code choice and signal processing techniques \cite{van2013almost}. 
For example, the effect of Doppler can be addressed by decreasing the code duration $T_d$, as shown in Fig. \ref{fig:mod_schemes}b \cite{roos2019radar}. 
Although PMCW transceivers require higher-bandwidth ADCs, as increased attention is being paid to 60 GHz communication, suitable ADCs in complementary metal-oxide semiconductor (CMOS) are becoming increasingly viable \cite{bourdoux2016PMCW}. 
Similar to FMCW signaling, the range resolution $\delta_R$ for PMCW radar is dependent on the system bandwidth $B$. 
The maximum unambiguous range $R_\text{max}$ and maximum resolvable velocity $v_\text{max}$ are dependent on the code duration $T_d$, and the total illumination time $T_I = N_\text{code} T_d$ for $N_\text{code}$ codes determines the Doppler resolution $\delta_v$ \cite{hakobyan2019high}. 

Another digital radar technique employs orthogonal frequency-division multiplexing (OFDM). 
Multicarrier OFDM has been employed for wideband communication for decades but has only recently been applied for radar signaling \cite{zhang2015ofdm}. 
The transmitted signal consists of a parallel transmission of coded signals at multiple subcarrier frequencies. 
Each subcarrier signal is modulated with a unique code, and a cyclic prefix (CP) of length $T_{cp}$ is transmitted before the symbol to prevent intersymbol interference and convert the channel convolution from linear to cyclical. 
For a symbol length of $T_\text{sym}$, $N_{sc}$ subcarriers are equally spaced with $\Delta_f = 1/T_\text{sym}$, which results in a total bandwidth of $B = N_{sc} \Delta_f$, as shown in Fig. \ref{fig:mod_schemes}c.
Due to the reciprocal relationship between the subcarrier spacing $\Delta_f$ and symbol duration $T_\text{sym}$, the multicarrier OFDM signal will not suffer from intercarrier interference because the sinc zero spacing occurs at the peak of each subcarrier signal \cite{roos2019radar}. 
This also allows for transmit orthogonality using spectral interleaving such that multiple transmitters are operated simultaneously using distinct subcarriers \cite{wang2012interleaved}. 
However, because of its large baseband bandwidth, OFDM radar requires high-bandwidth filtering and ADC hardware at the receiver. 
Recent research has used a stepped OFDM approach \cite{schweizer2018stepped}, which sweeps across a larger bandwidth with multiple steps, in attempt to address this downside. 
Alternatively, \cite{schindler2018mimo} combined a chirp signal with OFDM to increase the effective bandwidth, and \cite{knill2018high} employs a compressed-sensing approach to reduce sampling rates by randomly occupying smaller portions of the entire bandwidth. 
Again, the system bandwidth $B$ determines the range resolution $\delta_R$ for OFDM, and the Doppler resolution $\delta_v$ varies with the total transmission time $T_T = T_\text{sym} N_\text{sym}$ for $N_\text{sym}$ symbols. 
The maximum unambiguous range is determined by the length of the CP $T_{cp}$ and the maximum resolvable velocity is dependent on the pulse repetition frequency $f_\text{PRF} = 1/T_r$.  

Waveform design for SAR imaging is important for enabling the next generation of mmWave and THz sensors. 
An increasing push towards simultaneous communication and sensing, particularly in the THz regime, necessitates a thorough investigation into the optimization of signaling techniques \cite{chaccour2022seven,sarieddeen2020next,li2021Integrated}. 
Furthermore, with the emergence of reconfigurable intelligent surface (RIS)-aided systems and cognitive radio (CR), additional flexibility allows for improved waveform selection and coordination \cite{munochiveyi2021reconfigurable}. 

\subsection{Near-Field SAR Scanning Geometries}
\label{subsec:sar_scanning_modes}

Over the past several decades, mmWave and THz algorithms have been developed using various signaling techniques for high-resolution near-field SAR imaging. 
Here, we provide an overview of the common near-field SAR geometries for typical scanning modes capable of generating 2-D and 3-D images using the algorithms discussed later. 

\begin{figure}[ht]
\centering
    \begin{tabular}{c}
     \centering
     \includegraphics[width=0.3\textwidth]{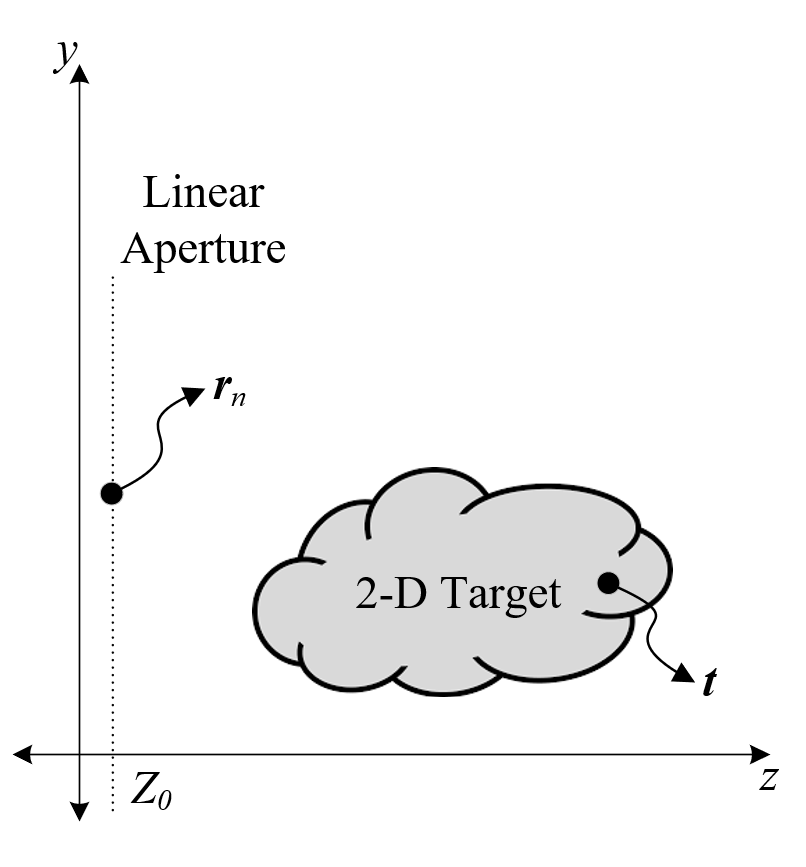} 
        \small(a)
    \end{tabular}
    \begin{tabular}{c}
     \centering
     \includegraphics[width=0.3\textwidth]{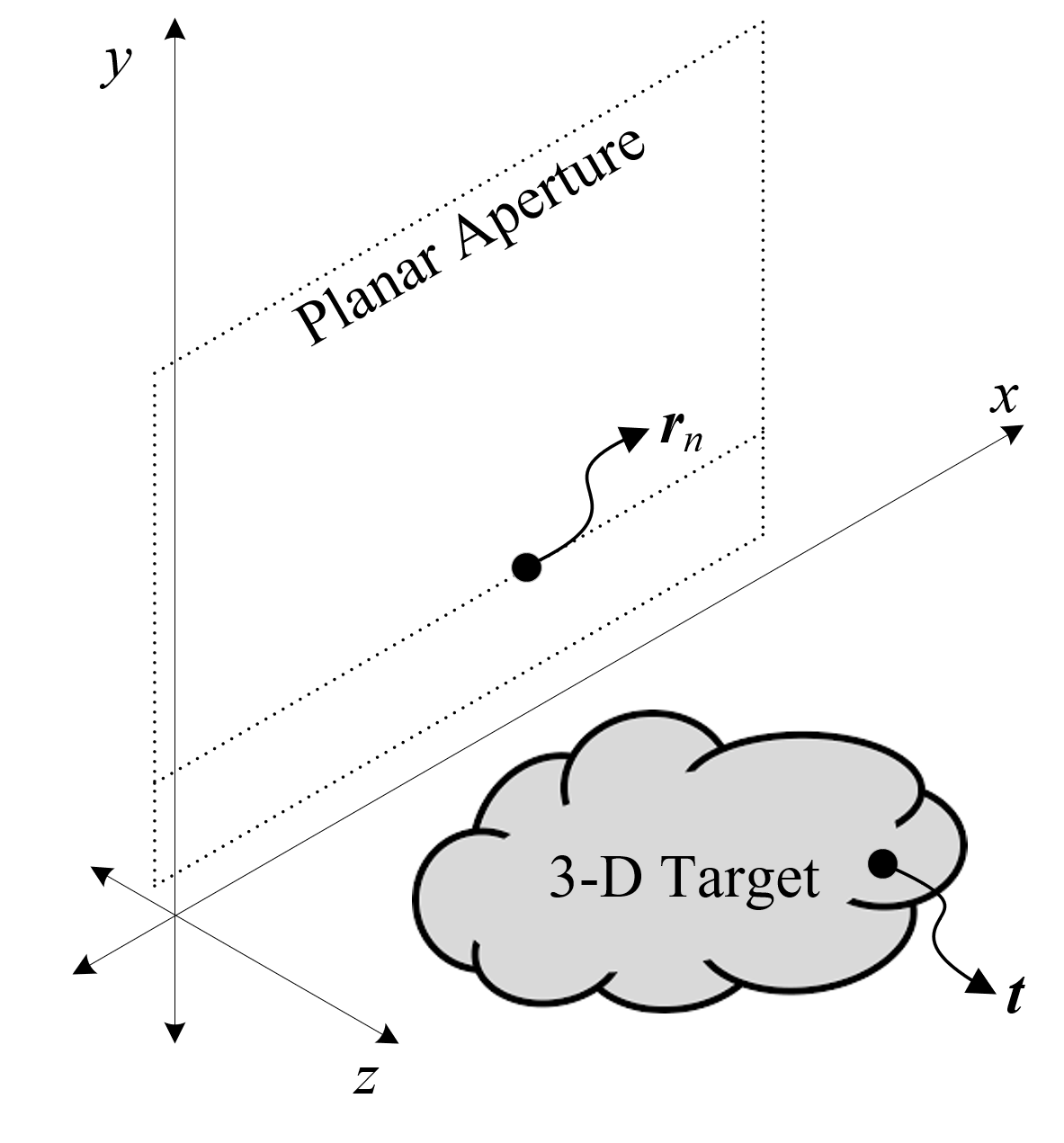} 
        \small(b)
    \end{tabular}
\caption{Geometry and coordinate definitions for SAR scanning regimes following the rectilinear scanning pattern in 1-D or 2-D. (a) Linear SAR. (b) Planar SAR.}
\label{fig:rectilinear_system_geometries}
\end{figure}

The rectilinear family of SAR scanning geometries comprises linear and planar synthetic apertures, as shown in Fig. \ref{fig:rectilinear_system_geometries}. 
For the simple 1-D linear aperture shown in Fig. \ref{fig:rectilinear_system_geometries}a, the radar transceivers are located at $\mathbf{r}_n = (y_n',Z_0)$ in the $y$-$z$ plane. 
Imaging of a 2-D continuously distributed target located at points $\mathbf{t} = (y,z)$ with reflectivity $\sigma(\mathbf{t})$ is achieved by illuminating the target across a wide bandwidth \cite{soumekh1998wide}. 
Near-field linear SAR imaging, similar to traditional strip-map SAR \cite{soumekh1999synthetic}, requires only a single pass to synthesize the linear aperture; however, images are limited to 2-D: vertical cross-range ($y$) and range ($z$). 
Hence, linear SAR has been applied to simple target scenarios, such as single-hand tracking \cite{smith2021An} and few-target subsurface imaging \cite{maisto2021sensor}. 

Planar SAR, shown in Fig. \ref{fig:rectilinear_system_geometries}b, is the 2-D extension of linear SAR, where a planar aperture is synthesized at the $z=Z_0$ plane in $x$-$y$-$z$ space. 
Hence, the transceiver locations are given by $\mathbf{r}_n = (x_n',y_n',Z_0)$ and the 3-D target is located at $\mathbf{t} = (x,y,z)$. 
Planar SAR has been applied for concealed object detection for several decades and has received increased attention as mmWave and THz hardware has become more affordable \cite{sheen2001three,lopez20003}. 
Previous studies have investigated synthesizing planar SAR geometries using either small-form-factor radars scanning in a raster pattern \cite{yanik2018millimeter,yanik2019cascaded,yanik2019near,wang2020csrnet,wang2021tpssiNet,smith2022ffh_vit} or 1-D scanning of linear multiple-input multiple-output (MIMO) arrays \cite{fan2020linearMIMOArbitraryTopologies,gao2018_1D_MIMO} to generate 2-D MIMO-SAR apertures. 
Although 1-D scanning of a large linear MIMO array provides more efficient scanning, raster scanning can achieve large apertures using inexpensive commercially-available mmWave radars \cite{yanik2020development,mohammadian2019sar}. 
As 5G and 6G hardware becomes increasingly prevalent, array imaging systems with mmWave and THz devices using simple scanning geometries such as linear and planar apertures are likely continue to increase in popularity for emerging tasks in sensing and robotics \cite{batra2021short,li2021Integrated,bourdoux2016PMCW,she2021tutorial,rappaport2018wireless,hong2021radar}. 

\begin{figure}[ht]
\centering
    \begin{tabular}{c}
     \centering
     \includegraphics[width=0.3\textwidth]{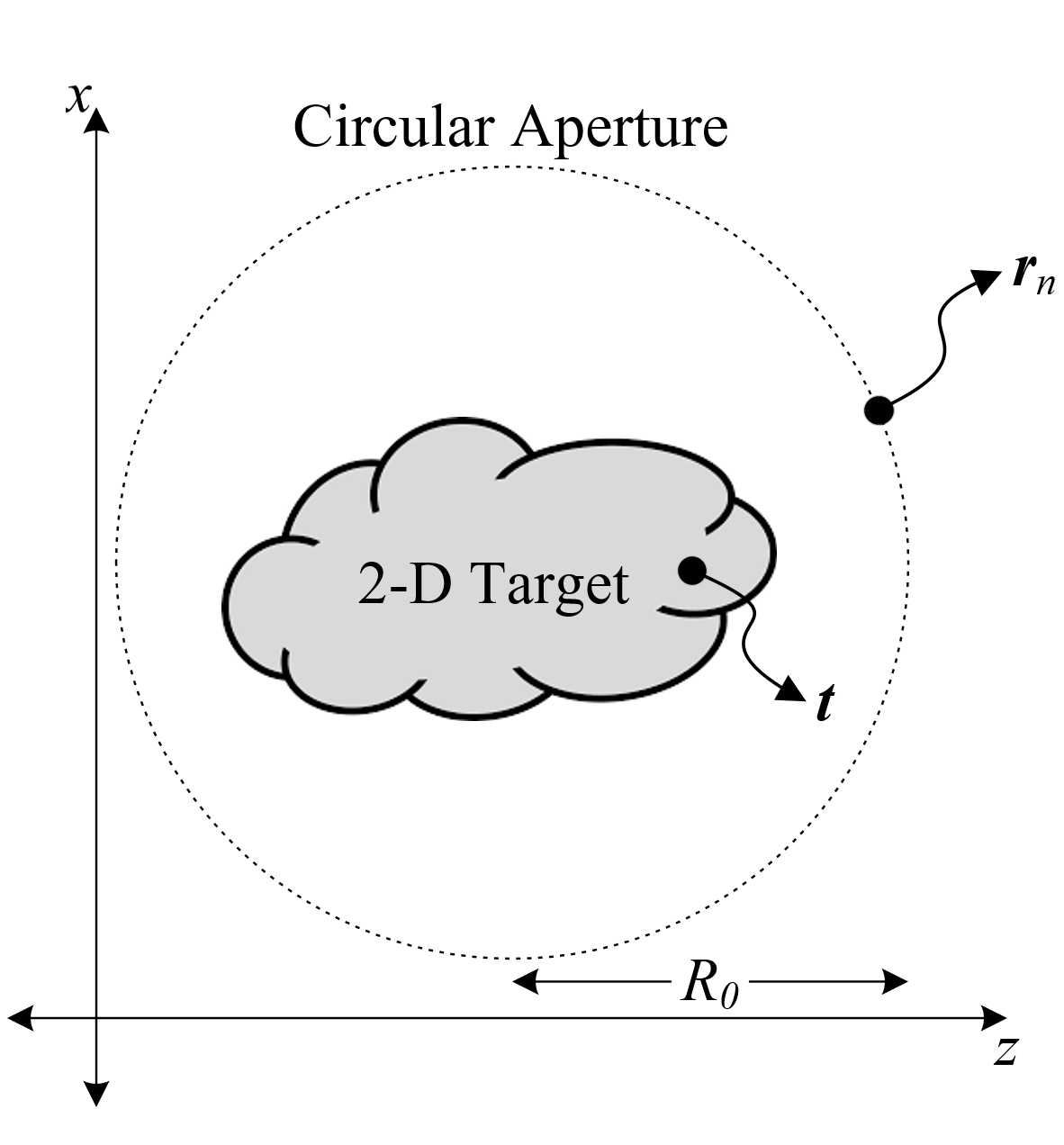} 
        \small(a)
    \end{tabular}
    \begin{tabular}{c}
     \centering
     \includegraphics[width=0.3\textwidth]{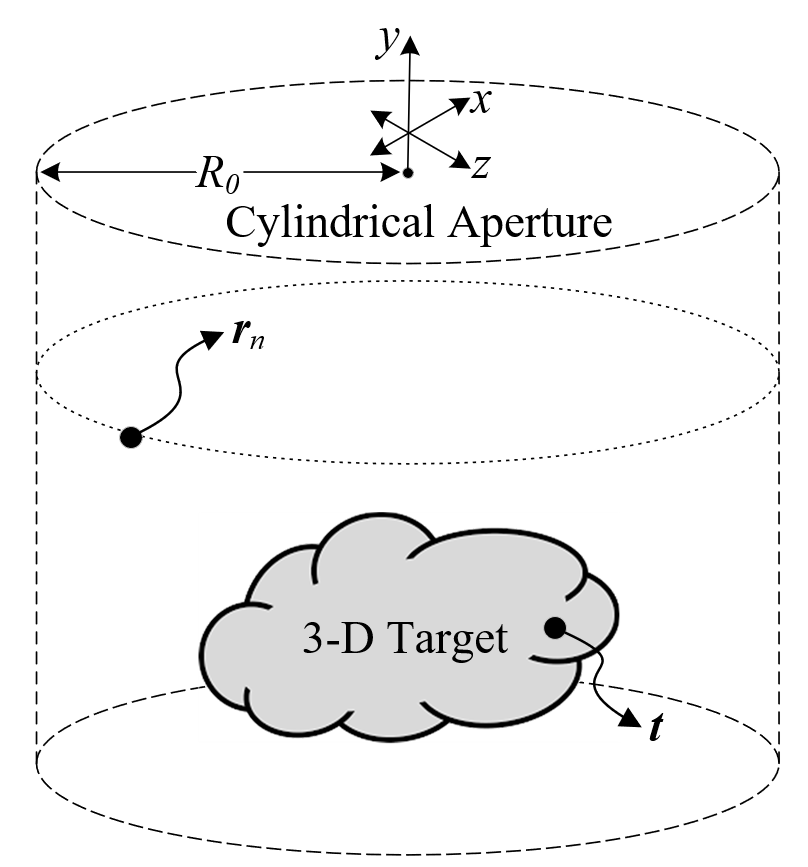} 
        \small(b)
    \end{tabular}
\caption{Geometry and coordinate definitions for SAR scanning regimes following the polar scanning pattern in 1-D or 2-D. (a) Circular SAR. (b) Cylindrical SAR.}
\label{fig:polar_system_geometries}
\end{figure}

The polar family of near-field SAR scanning regimes comprises circular and cylindrical apertures, as shown in Fig. \ref{fig:polar_system_geometries}. 
Circular SAR (CSAR) involves scanning a transceiver around the target or rotating the target at a fixed radial distance $R_0$ from the stationary radar, as shown in Fig. \ref{fig:polar_system_geometries}a. 
Hence, the array elements are located at $\mathbf{r}_n = (R_0 \cos \theta_n, R_0 \sin \theta_n)$ in $y$-$z$ space, where $\theta_n$ is the angle of rotation and the target is located at points $\mathbf{t} = (y,z)$. 
Similar to linear SAR, imaging in the CSAR regime is confined to two dimensions \cite{gao2016efficient,demirci2011back}. 
However, CSAR can provide superior robustness compared to linear SAR, as the target is illuminated from multiple perspectives across the circular aperture, thereby overcoming occlusion for certain scenarios wherein the target plane is known \cite{smith2020nearfieldisar}. 

Cylindrical SAR, also known as elevation-CSAR or ECSAR \cite{bryant20033}, extends CSAR in the elevation direction to achieve a cylindrical synthetic aperture geometry, as shown in Fig. \ref{fig:polar_system_geometries}b. 
The radar transceivers are located at $\mathbf{r}_n = (R_0 \cos \theta_n, y_n', R_0 \sin \theta_n)$ in $x$-$y$-$z$ space along with the target at locations $\mathbf{t} = (x,y,z)$. 
Cylindrical SAR apertures can be synthesized by several means, including cylindrical raster patterns \cite{smith2020nearfieldisar}, a linear MIMO array along the $y$-direction scanning radially \cite{detlefsen2005effective,gao2018cylindricalMIMO}, or a circular MIMO array scanned linearly along the $y$-direction \cite{berland2021cylindricalMIMOSAR}. 
The most common approach is to use a linear MIMO array and either rotate the object under test or rotate the array around the target or human subject \cite{detlefsen2005effective}. 
Similarly, for imaging scenarios with complex target scenes, such as concealed item detection, where occlusion may degrade detection performance, cylindrical SAR is advantageous over planar SAR for 3-D imaging \cite{smith2020nearfieldisar}. 

Alternatively, with the Internet of Things (IoT), applications necessitating irregular scanning geometries have gained attention in recent studies \cite{alvarez2019freehand,smith2022ffh_vit,vasileiou2022efficient}. 
Freehand SAR imaging, wherein a handheld compact radar is moved throughout space by a user to synthesize an aperture, is of particular interest for enabling high-resolution sensing in edge applications \cite{alvarez2021towards,smith2022efficient}. 
However, two primary challenges exist in irregular SAR imaging tasks. 
First, high-fidelity position estimation is required to localize the transceiver at each sample location of the synthetic aperture. 
Infrared (IR) camera tracking systems \cite{alvarez2019freehand} and compact stereo camera sensors \cite{alvarez2021towards} have been investigated in prior work and demonstrate sufficient position estimation for medium-fidelity imaging. 
However, smaller THz wavelengths will pose a new challenge for localization and will require additional system- and signal-level innovations.  
The second challenge is efficient image recovery on computationally constrained hardware. 
Performing image reconstruction with irregular array geometries that do not conform to the typical SAR scanning patterns in Figs. \ref{fig:rectilinear_system_geometries} and \ref{fig:polar_system_geometries} for IoT and edge applications demands further research into flexible and highly efficiency imaging algorithms. 
In \cite{smith2022efficient}, an efficient algorithm is proposed assuming a semi-planar geometry; however, additional investigations into sparse solutions and efficient backprojection may be necessary for practical commercial implementation. 

\subsubsection{Raw SAR Data}
\label{subsubsec:raw_data}
For each of the SAR scanning geometries discussed, the radar data at each transceiver can be expressed as the superposition of the reflected signals from the distributed target at locations $\mathbf{t}$. 
Throughout this paper, a monostatic full-duplex assumption is placed on the transceivers, although the signal model and reconstruction algorithms herein can easily be extended to a multistatic MIMO scenario, as in \cite{gao2018_1D_MIMO,yanik2020development,berland2021cylindricalMIMOSAR}. 
Without loss of generality, assuming the Born approximation and isotropic propagation, the raw SAR signal at the $\mathbf{r}_n$ radar element can be expressed as
\begin{equation}
    \label{eq:sar_data}
    s(\mathbf{r}_n,f) = \int_{\mathbf{V}_\sigma} \frac{\sigma(\mathbf{t})}{||\mathbf{t} - \mathbf{r}_n||^2 } e^{-j \frac{4 \pi f}{c} ||\mathbf{t} - \mathbf{r}_n||} d\mathbf{t}, 
\end{equation}
where $\mathbf{t}$ denotes the locations of the target, which occupies the volume $\mathbf{V}_\sigma$ with the continuous reflectivity function $\sigma(\mathbf{t})$ describing the wave reflected from the scatterer at $\mathbf{t}$, the frequency points in the frequency range of the system are given by $f$, $||\cdot||$ is the Euclidean 2-norm, and $c$ is the speed of light. 
The goal of SAR imaging is to recover the reflectivity image of the target $\sigma(\mathbf{t})$ by inverting (\ref{eq:sar_data}). 
A variety of approaches have been explored to recover the reflectivity function over recent decades offering unique advantages including numerical accuracy, efficiency, and simplicity.

\subsection{Fundamentals of Near-Field SAR Imaging Algorithms}
\label{subsec:sar_imaging_algorithms}
In this section, we briefly introduce several common SAR image reconstruction algorithms for the previously detailed geometries. 
Each of the algorithms is implemented, and source code is made available in the interactive toolbox to improve the accessibility of near-field SAR imaging concepts, accelerate algorithm development, and improve research on data-driven SAR approaches.  

\subsubsection{Backprojection Algorithm}
\label{subsubsec:bpa}
The direct solution to (\ref{eq:sar_data}) is the migration-based imaging method known as the backprojection algorithm or BPA. 
The BPA is flexible for any array configuration and can be directly applied to all four scanning regimes described in the previous section. 
However, the BPA requires computing a point-by-point matched filter response for every sample point and voxel in the reconstructed volume, as given by
\begin{equation}
    \label{eq:bpa}
    \hat{\sigma}(\Tilde{\mathbf{t}}) \approx \iint s(\mathbf{r}_n,f) e^{j \frac{4 \pi f}{c} ||\Tilde{\mathbf{t}} - \mathbf{r}_n||} d\mathbf{r}_n df, 
\end{equation}
where $\Tilde{\mathbf{t}}$ is the set of voxels selected by the user in the spatial region over which the matched filtering operation is performed. 
The result is an image with dimensionality of $\Tilde{\mathbf{t}}$ comprising intensity values corresponding to the estimated reflectivity of the target scene at each voxel location. 
BPA imaging algorithms have been applied to mmWave \cite{yanik2020development} and THz \cite{jaeschke20143d,batra2021Submm} imaging for concealed item detection and non-destructive testing. 
However, computational inefficiency is considerable issue, particularly for high-resolution 3-D imaging tasks \cite{smith2022efficient,gao2018_1D_MIMO}. 
Prior efforts have attempted to improve the efficiency of the BPA by exploiting convolution relations in (\ref{eq:bpa}) to reduce complexity using Fourier analysis \cite{yanik2020development,gao2016efficient,yanik2018millimeter}, using fast-factorized BPA \cite{moll2012towards}, or implementation on a field programmable gate array (FPGA) \cite{batra2021fpga}. 
A vectorized implementation of the BPA is included in the interactive toolbox developed in this study, which leverages GPU parallelization to improve computational efficiency. However, for many sensing applications, the BPA is considered computationally intractable. 

\subsubsection{Rectilinear Range Migration Algorithm}
\label{subsubsec:rma}
Alternatives to the BPA have quickly emerged, leveraging the FFT to drastically reduce the computational complexity while placing constraints on the SAR geometry. 
The main challenge in inverting (\ref{eq:sar_data}) is the spherical wavefront that must be considered for near-field SAR, which is realized by the nonlinear phase term. 
Based on the scanning geometry, the phase term can be decomposed using the method of stationary phase (MSP) to be represented as a superposition of planewaves in the spatial spectral domain \cite{papoulis1968systems}. 
After the MSP is applied, the BPA integral in (\ref{eq:bpa}) can be simplified using Fourier analysis, where the spherical wavefront can be addressed using either Stolt interpolation \cite{lopez20003} or non-uniform FFT (NUFFT) approaches \cite{gao2016efficient,gao2018_1D_MIMO,fan2020linearMIMOArbitraryTopologies,smith2022efficient}. 

First, we examine the planar SAR geometry, given in Fig. \ref{fig:rectilinear_system_geometries}b, where $\mathbf{r}_n = (x_n',y_n',Z_0)$ and $\mathbf{t} = (x,y,z)$, such that the target is centered at the origin. 
Hence, the MSP can be applied to phase term in (\ref{eq:sar_data}), yielding
\begin{equation}
    \label{eq:planar_phase}
    e^{-j \frac{4 \pi f}{c} ||\mathbf{t} - \mathbf{r}_n||} = e^{-j \frac{4 \pi f}{c} \sqrt{(x - x_n')^2 + (y - y_n')^2 + (z - Z_0)^2}},
    \approx \iint e^{-j(k_x(x - x_n') + k_y(y - y_n') + k_z(z - Z_0))} d k_x d k_y,
\end{equation}
where $k_x$, $k_y$, and $k_z$ are the spatial spectral variables along the $x$-, $y$-, and $z$-directions, respectively, satisfying the relation
\begin{equation}
    \label{eq:wavenumber_relation}
    k_z^2 = 4k^2 - k_x^2 - k_y^2, \quad k_x^2 + k_y^2 \leq 4k^2,
\end{equation}
with $k \triangleq 2 \pi f / c$ being the radial wavenumber corresponding to the frequency $f$. 
Substituting (\ref{eq:planar_phase}) into (\ref{eq:sar_data}) and using the Fourier transform relations yields \cite{yanik2020development}
\begin{equation}
    S(k_x,k_y,k) = \sigma(k_x,k_y,k_z)e^{j k_z Z_0}. 
\end{equation}
Hence, the target image can be recovered by performing a 3-D inverse FFT (IFFT) of $\sigma(k_x,k_y,k_z)$ as
However, the SAR data spectrum ${S}(k_x,k_y,k)$ is sampled along the radial $k$-domain and must be accounted for prior to the IFFT using either Stolt interpolation \cite{lopez20003} or a NUFFT approach \cite{gao2018_1D_MIMO}. 
This technique, known as the range migration algorithm (RMA), reduces the complexity of the BPA from approximately $O(N^6)$ to $O(N^3 \log N)$ \cite{sheen2001three,lopez20003,yanik2020development} for 3-D imaging using planar SAR.
The planar RMA can be summarized as follows
\begin{equation}
    \label{eq:planar_rma}
    \sigma(x,y,z) = \text{IFT}_{\text{3D}}^{(k_x,k_y,k_z)} \bigg[ \mathcal{S}_R \left[ S(k_x,k_y,k_z) e^{-j k_z Z_0} \right] \bigg],
\end{equation}
where $\mathcal{S}_R[\cdot]$ is the rectilinear Stolt interpolation operator from $k$ to $k_z$ sampling. 
Similarly, for linear SAR, (\ref{eq:planar_rma}) can be simplified for 1-D scanning and a 2-D image scene. 

\begin{figure}[ht]
    \centering
    \includegraphics[width=0.55\textwidth]{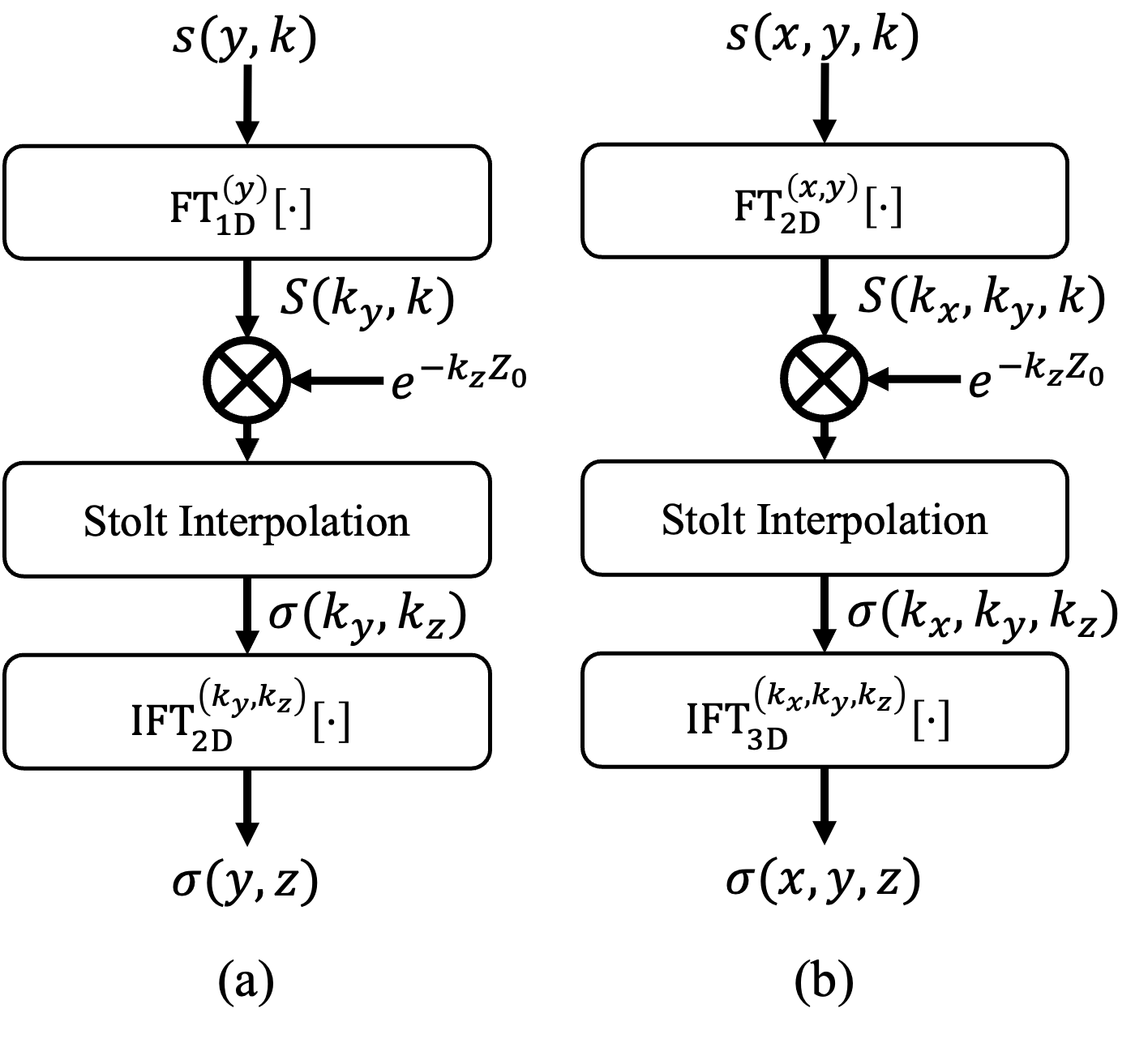}
    \caption{Flowcharts of RMA image reconstruction algorithms for (a) linear and (b) planar SAR scanning modes shown in Fig. \ref{fig:rectilinear_system_geometries}.}
    \label{fig:rect_sar_algorithm_flowcharts}
\end{figure}

The RMA algorithm for the rectilinear geometries, planar and linear, is detailed in Fig. \ref{fig:rect_sar_algorithm_flowcharts}. 
For both scanning patterns, the majority of the algorithm is performed in the spatial spectral domain, or $k$-space, and the image is recovered in the final step using an $N$-D IFFT. 
As explored in \cite{yanik2019sparse,wang2018wavenumber}, the non-zero area of the $k$-space is dependent on the bandwidth and aperture size. 
This is a key insight into the derivation of the spatial resolution of the rectilinear SAR imaging, which is bounded by the sinc-function phenomenon. 
Limited bandwidth and spatial sampling (synthetic aperture size), effectively apply a windowing function that imposes a sinc-effect in the resulting image. 
The bandwidth limitation phenomenon was leveraged in \cite{smith2023dual_radar} by approaching the super-resolution problem as a joint extrapolation and super-resolution problem, demonstrating superior performance than an algorithm operating exclusively in one domain or the other. 
For similar SAR super-resolution and image processing applications, understanding this underlying mechanism allows for more advanced algorithms that operate on a combination of the raw radar signal, reconstructed image, and intermediate reconstructions ($k$-space), as discussed in Section \ref{subsec:signal_domain_vs_image_domain}. 

The resolution for the planar RMA along each direction in 3-D space can be expressed as
\begin{align}
\label{eq:planar_spatial_resolution}
\begin{split}
    \delta_x &= \frac{\lambda_c Z_\text{ref}}{2 D_x}, \\
    \delta_y &= \frac{\lambda_c Z_\text{ref}}{2 D_y}, \\
    \delta_z &= \frac{c}{2B},
\end{split}
\end{align}
where $\lambda_c$ is the center frequency, $Z_\text{ref}$ is the distance from the aperture to a reference point on the target, and aperture sizes along the $x$- and $y$-directions are given by $D_x$ and $D_y$, respectively. 
The RMA, also known as the $\omega$-$k$ or $\omega$-$f$ algorithm, is widely used in the literature \cite{batra2021Submm,yanik2019cascaded,yanik2020development,lopez20003,sheen2001three,smith2022efficient,smith2022ffh_vit} and is included in the open-source toolbox developed in this study. 

\subsubsection{Polar Range Migration Algorithm}
\label{subsubsec:pfa}
Similarly, the RMA can be applied to circular and cylindrical synthetic apertures following a similar procedure. 
The curved aperture RMA is also referred to as the Polar Formatting Algorithm (PFA) \cite{gezimati2022circular,gezimati2023curved}. 
Following the geometry in \ref{fig:polar_system_geometries}b with $\mathbf{r}_n = (x_n', y_n', z_n')$ and $\mathbf{t} = (x,y,z)$, where the target is centered at the origin. 
Substituting $x_n' = R_0 \cos \theta_n$ and $z_n' = R_0 \sin \theta_n$ to satisfy the cylindrical SAR geometry and using (\ref{eq:wavenumber_relation}), the relationships between the rectangular and cylindrical spatial wavenumbers are expressed as
\begin{align}
    \label{eq:cylindrical_wavenumbers}
    \begin{split}
        k_x &= k_r \cos \alpha, \\
        k_z &= k_r \sin \alpha, \\
        k_r^2 &= k_x^2 + k_z^2 = 4k^2 - k_y^2,
    \end{split}
\end{align}
where the relation in (\ref{eq:wavenumber_relation}) holds for the rectangular spatial spectral domain. 

Using this geometry, the MSP can be applied to the phase term in (\ref{eq:sar_data}) to yield
\begin{equation}
    \label{eq:cylindrical_phase}
    e^{-j \frac{4 \pi f}{c} ||\mathbf{t} - \mathbf{r}_n||} = e^{-j \frac{4 \pi f}{c} \sqrt{(x - x_n')^2 + (y - y_n')^2 + (z - z')^2}}
    \approx \iiint e^{-j(k_x(x - x_n') + k_y(y - y_n') + k_z(z - z_n'))} d k_x d k_y d k_z.
\end{equation}
Substituting (\ref{eq:cylindrical_phase}) into (\ref{eq:sar_data}) and applying the Fourier transform relations results in 
\begin{equation}
    \label{eq:circular_conv_step}
    S(\theta, k_y, k) = \int \left[ \int \sigma(\alpha, k_y, k_r) e^{-j k_r R_0 \cos (\theta - \alpha)} d \theta \right] k_r d k_r,
\end{equation}
where the term inside the brackets is a convolution integral in the $\theta$ domain and $\theta$ and $\alpha$ are coincident. 
Hence, taking the Fourier transform with respect to $\theta$ leads to
\begin{equation}
    \label{eq:conv_ft}
    S(k_\theta, k_y, k) = \int \sigma(k_\alpha, k_y, k) G(k_\theta, k_y, k) k_r d k_r,
\end{equation}
where
\begin{equation}
    \label{eq:azimuth_focusing}
    G(k_\theta, k_y, k) \triangleq \text{FT}_{\text{1D}}^{(\theta)} \left[ e^{-j \sqrt{4k^2 - k_y^2} R_0 \cos \theta} \right].
\end{equation}
Since the integral in (\ref{eq:azimuth_focusing}) consists of the values of the object on the Ewald sphere given by (\ref{eq:wavenumber_relation}) and (\ref{eq:cylindrical_wavenumbers}), imposing a $\delta$-function behavior with respect to $k_r$ \cite{detlefsen2005effective}, the inversion can be simplified to
\begin{equation}
    \label{eq:cylindrical_inversion}
    \sigma(k_\theta, k_y, k_r) = S(k_\theta, k_y, k) G^*(k_\theta, k_y, k),
\end{equation}
where $[\cdot]^*$ is the complex conjugation operation from which the spatial spectral reflectivity function can be recovered using an inverse Fourier transform across $k_\theta$ as
\begin{equation}
    \label{eq:cylindrical_inverse_ft}
    \sigma(\theta, k_y, k_r) = \text{IFT}_\text{1D}^{(k_\theta)} \left[ S(k_\theta, k_y, k) G^*(k_\theta, k_y, k) \right]. 
\end{equation}

Finally, the reflectivity function can be recovered from (\ref{eq:cylindrical_inverse_ft}) using Stolt interpolation from the cylindrical coordinates to rectangular spatial spectral coordinates, followed by a 3-D IFFT \cite{smith2020nearfieldisar} or using an NUFFT approach \cite{gao2016efficient,gao2018cylindricalMIMO}.
The final step of the cylindrical RMA can be written as
\begin{equation}
    \label{eq:cylindrical_rma}
    \sigma(x,y,z) = \text{IFT}_{\text{3D}}^{(k_x,k_y,k_z)} \bigg[ \mathcal{S}_P \left[ \sigma(\theta, k_y, k_r) \right] \bigg],
\end{equation} 
where $\mathcal{S}_P[\cdot]$ is the polar Stolt interpolation operator to a uniform rectangular spatial spectral $k_x$-$k_y$-$k_z$ grid. 
For circular SAR, (\ref{eq:cylindrical_rma}) can be simplified to ignore the height of the cylinder along the $y$-direction, as in \cite{gao2016efficient,gezimati2022circular}. 

\begin{figure}[ht]
    \centering
    \includegraphics[width=0.55\textwidth]{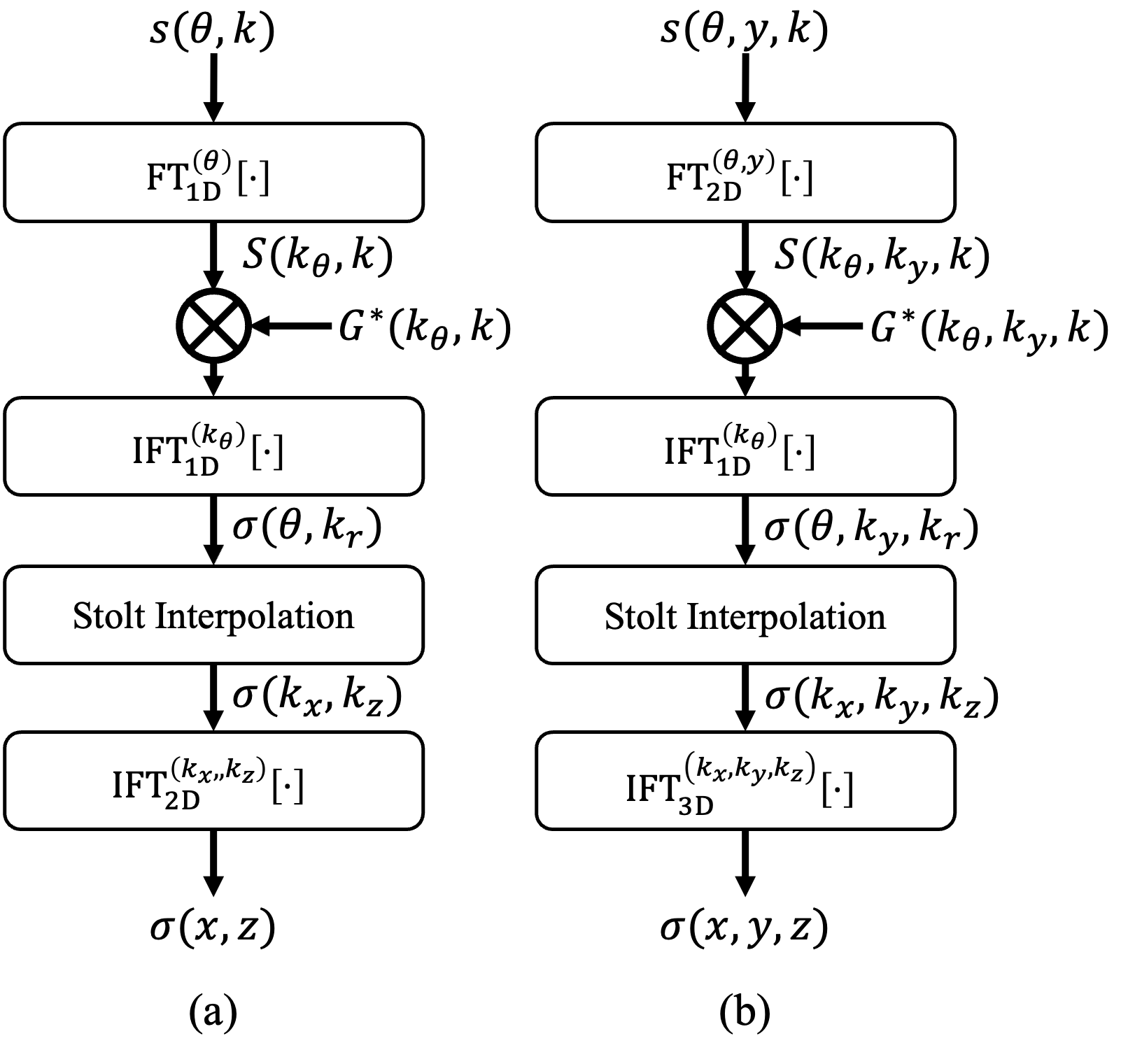}
    \caption{Flowcharts of RMA image reconstruction algorithms for (a) circular and (b) cylindrical SAR scanning modes shown in Fig. \ref{fig:polar_system_geometries}.}
    \label{fig:polar_sar_algorithm_flowcharts}
\end{figure}

Similar to the rectilinear RMA reconstruction algorithms, the polar RMA for cylindrical and circular geometries, shown in Fig. \ref{fig:polar_sar_algorithm_flowcharts}, operate primarily in the spatial spectral domain. 
Again, the $k$-space occupied by the target is directly proportional to the aperture size and system bandwidth \cite{gao2016efficient} and impacts the reconstructed image resolution. 
However, along the radial direction, the resolution is dictated by the behavior of the first-order Bessel function \cite{gao2018cylindricalMIMO} as opposed to the sinc-function. 

The resolution for the cylindrical RMA along the vertical direction $y$ and radial direction $\rho$ can be expressed as
\begin{align}
\label{eq:cylindrical_spatial_resolution}
\begin{split}
    \delta_y &= \frac{\lambda_c R_0}{2 D_y}, \\
    \delta_\rho &= \frac{2.4}{k_\text{max} + k_\text{min}},
\end{split}
\end{align}
where $k_\text{max}$ and $k_\text{min}$ are the maximum and minimum wavenumbers, respectively, obtained by analysis of the point spread function (PSF) Bessel function \cite{gao2018cylindricalMIMO,smith2020nearfieldisar}. 
Although this analysis focuses on monostatic apertures \cite{detlefsen2005effective}, variants of the RMA for MIMO cylindrical SAR have been developed for several MIMO scanning modes \cite{smith2020nearfieldisar,gao2018cylindricalMIMO,berland2021cylindricalMIMOSAR}. 
The cylindrical RMA is implemented in the included software platform, allowing rapid prototyping of systems in addition to dataset generation, which is a crucial issue for emerging data-driven SAR algorithms. 

Other popular algorithms for near-field imaging using common SAR geometries include the chirp scaling algorithm (CSA) \cite{wang2020AnEfficientCSA}, which provides improved efficiency over the RMA at the cost of reduced numerical accuracy, and the phase-shift migration (PSM) algorithm \cite{gao2019implementation,gao2020study}, which employs the exploding field reflector model to yield improved accuracy compared with the RMA. 
Alternatively, algorithms have been presented to mitigate the interpolation or NUFFT requirements of the RMA \cite{yanik2020development,li2013interpolation}. 

As data-driven algorithms have begun to dominate the fields of computer vision and image processing over the past few decades, a similar trend is emerging for near-field SAR imaging. 
With the increase in available parallelized computational power, machine learning and deep learning algorithms for SAR image processing, restoration, super-resolution, segmentation, and object detection are being actively developed. 
However, a sufficient understanding of the underlying mechanics of the SAR imaging problem is essential for developing learning-based algorithms that optimally leverage the various aspects of the imaging data at each step of the image reconstruction process. 
For example, understanding the spectral properties of the signal at each step of the reconstruction process and developing a rigorous intuition for the relationships between aperture size and cross-range resolution provides considerable insight for developing image processing, classification, and super-resolution algorithms. 

\section{Radar Imaging Systems from GHz to THz}
\label{sec:sar_systems}
In this section, we briefly review recent developments in array imaging systems spanning mmWave and THz spectra. 
With the emergence of 5G, 6G, and IoT technologies, mmWave and THz hardware are drawing increasing attention from research and industry. 
As UWB hardware becomes more available, EM sensing using mmWave and THz waves is becoming increasingly vital on a wide array of platforms for many sensing applications. 

\subsection{Emerging Systems for THz Sensing Applications}
\label{subsec:emerging_systems}
A promising aspect of emerging 5G and 6G protocols is the incorporation of high-resolution, high-throughput joint communications and sensing (JCS) \cite{chaccour2022seven}. 
JCS entails the synchronization of communication and sensing protocols, providing an elegant solution to spectrum sharing in mmWave and THz wavelengths if adequate coordinate can be achieved. 
Applications spanning augmented reality (AR), virtual reality (VR), extended reality (XR), Industry 4.0, and Digital Twins have been explored leveraging JCS for high-resolution localization and imaging on computationally constrained platforms \cite{sarieddeen2020next}. 
In \cite{li2021Integrated}, a JCS technique was proposed for joint communications and freehand imaging similar to previous radar-focused efforts \cite{smith2022efficient,alvarez2021freehand}. 
JCS has been proposed for a variety of unique network applications. 
Vehicular mmWave and THz networks using JCS were explored in \cite{petrov2019on} leveraging high-resolution sensing for understanding of dynamic environments and creating large-bandwidth communication channels. 
In \cite{aliaga2023enhancing} a technique was introduced for improving CubeSat networks using orbital angular momentum. 
Similar JCS systems have been investigated for synchronizing communication and sensing capabilities on UAV \cite{chang2022integrated} and low earth orbit (LEO) satellite networks \cite{aliaga2022joint}. 
JCS holds significant potential to impact several domains by integrating communication and sensing functionalities, thereby improving performance and enabling novel technologies. 

THz and mmWave sensing has also recently attracted interest for IoT applications. 
In \cite{hong2021radar}, JCS is explored for radar-communications integration to jointly leverage the advantages of massive MIMO communication arrays with micro-Doppler radar sensing. 
Medical imaging using IoT has also been investigated, with an emphasis on incorporating emerging data-driven algorithms \cite{banerjee2020medical}. 
By leveraging THz and mmWave sensing paired with IoT, medical imaging techniques can benefit from improved resolution and accuracy, enabling more prices diagnoses and treatment planning. 
Furthermore, THz IoT sensors have shown promise in the field of early gesture recognition \cite{min2021early}. 
This involves the rapid detection of hand gestures with minimal sensing time, which is crucial for applications such as human-computer interaction and gesture-based control systems. 
High-resolution mmWave and THz IoT imaging systems offer the capability to capture fine spatial signatures and subtle movements, enabling efficient and accurate input devices. 

Additionally, mmWave and THz systems emerged as viable solutions for non-line-of-sight (NLOS) sensing and imaging applications. 
Several studies have introduced systems and algorithms are NLOS imaging using mmWave SAR for high-resolution imaging of third-order reflections. 
This technique was extended in \cite{cui2022seeing} to create a full-body scanning system using THz hardware to reconstruct high-resolution images using reflections from drywall and an algorithm for reconstructing images from multiple reflections was introduced. 
Batra \textit{et al.} proposed systems to achieve long-range NLOS THz imaging of up to fifth-order reflections and with a range of greater than 6 m at a bandwidth of 110 GHz. 
In another study \cite{grossman2022passive}, researchers investigated a passive THz NLOS imaging system for localization, orientation, and pose estimation of human subjects. 
This system leverages THz sensing to capture subtle reflections and variations of the signal from a wall to reconstruct accurate images of the subjects without LOS to the passive device. 
These advancements in NLOS sensing and imaging systems enable various promising applications, including surveillance, search and rescue, and object recognition in challenging environments. 
However, they also create certain privacy and security risks for undesired remote sensing, which must be considered, and countermeasures should be explored further. 

\subsection{System Design Considerations for SAR Imaging Testbeds}
\label{subsec:sar_testbed_systems}
Implementation of array imaging systems is a crucial step in realizing high-resolution sensing with mmWave and THz hardware. 
Traditional short-range SAR testbed systems have relied on precise mechanical scanning lab equipment to create regular SAR geometries such as those in Figs. \ref{fig:rectilinear_system_geometries} and \ref{fig:polar_system_geometries}. 
Some systems have employed ball screw linear rails, driven by stepper drivers, due to their low cost, but suffer from slow scanning speeds, rendering applications such as full-human body scanning and real-time imaging infeasible \cite{yanik2018millimeter,yanik2019near,wei2021non,smith2020nearfieldisar,yanik2019sparse,wang2021rmistnet,wang2021tpssiNet}. 
Others have employed more expensive laboratory-grade ball screw linear rails to achieve faster scanning speeds \cite{yanik2019cascaded,saqueb2018THz,yanik2020development,gao2018_1D_MIMO,gao2018cylindricalMIMO,fan2020linearMIMOArbitraryTopologies}. 
Meanwhile, belt-driven rails have gained increasing popularity because of their ability to achieve higher speeds at comparatively inexpensive costs, rendering them more suitable for commercial applications \cite{smith2022efficient,smith2023dual_radar,vasileiou2022efficient,smith2022ffh_vit}. 
However, achieving rapid SAR scanning, regardless of geometry, requires overcoming several key system challenges. 

For slow-moving SAR applications, a constant velocity model can be applied, yielding a linear relationship between position and time.
Hence, to create an evenly-spaced synthetic aperture, the radar can be activated with constant periodicity across time. 
However, when operating high speeds, the scanner's acceleration can no longer be ignored, and the previous assumption does not hold. 
As a result, the radar must be synchronized to illuminate the target scene aperiodically across time to create a uniform grid of spatial samples. 
In \cite{yanik2020development}, a hardware synchronization platform is introduced to overcome this issue and enable high-speed scanning. 
The proposed technique is extended in \cite{smith2023dual_radar} to synchronize multiple radar systems to create useful synthetic apertures for multiband radar imaging. 
SAR can also play a vital role in understanding scenes for AR/VR applications. 
However, timing, spatial synchronization, and aperture construction are crucial to the success of a dynamic SAR scanning system. 
The synchronization techniques developed in \cite{yanik2020development,smith2023dual_radar,smith2022novel} paired with localization techniques, such as \cite{alvarez2019freehand,alvarez2021towards}, could extend traditional SAR to head-mounted or handheld applications. 

\subsection{Evolution of High-Bandwidth mmWave and THz Circuits and Imaging Systems}
Over the past decade, there has been remarkable progress in the development of mmWave and THz imaging hardware, enabling new possibilities in various applications. 
One notable advancement is the widespread adoption of commercial off-the-shelf (COTS) radars, with Texas Instruments' industrial 62 GHz and automotive 79 GHz radars, featuring 4 GHz bandwidth, gaining significant popularity \cite{TI:ramasubramanian2017mmwave}. 
These COTS radars are proven, versatile, and cost-effective solutions for a host of mmWave imaging and sensing tasks \cite{yanik2020development,wei2021non,smith2021sterile,smith2020nearfieldisar,vasileiou2022efficient,smith2023dual_radar}. 

In recent research, novel radar transceivers have been proposed to push the boundaries of imaging at higher frequencies and bandwidth. 
For instance, an 80 GHz SiGe-based radar transceiver with an impressive 25 GHz bandwidth was presented in \cite{pohl2018compact}. 
Gao \textit{et al.} showcased a vector network analyzer (VNA)-based THz SAR imaging system. 
This system operated at a center frequency of 100 GHz with a 15.75 GHz bandwidth \cite{gao2020study}. 
Additionally, a SiGe transceiver with a center frequency of 146 GHz and an impressive 48 GHz bandwidth was introduced in \cite{jaeschke2014high}. 
Building upon this work, the same authors proposed a monolithic microwave integrated circuit (MMIC) with a center frequency of 231 GHz and a 42 GHz bandwidth \cite{jaeschke20143d}. 

Furthermore, recent progress demonstrates the feasibility of high-resolution radar imaging systems as operating frequencies approach the THz range. 

In \cite{stanko2016millimeter}, a lower THz region imaging radar was proposed with 28.8 GHz bandwidth operating at 0.3 THz. 
Ding \textit{et al.} introduced a system operating at a center frequency of 0.54 THz and bandwidth of 51 GHz based on a lens beamsplitter design \cite{ding2013thz}. 
In 2008, an early demonstration of THz imaging radar boasted a center frequency of 0.6 THz with a bandwidth of 28.8 GHz \cite{cooper2008penetrating}.
The authors of \cite{saqueb2018THz} detailed a rail-based SAR imaging system operating at 0.625 THz with an impressive bandwidth of 0.25 THz.
Another notable development in THz imaging can be found in \cite{assou2023synthetic}, where a photoconductive antenna array was employed for 3-D imaging at a center frequency of 0.8 THz.
The system achieves a bandwidth of 0.4 GHz, facilitating high-resolution volumetric imaging and material characterization. 
Continuing the trend of pushing the boundaries of THz radar technology, recent work introduced an imaging system with a center frequency of 1.3 THz with a bandwidth of 0.4 THz \cite{batra2021Submm}. 
These systems represent a significant advancement in THz radar capabilities, allowing for even finer details to be resolved and expanding the potential applications of THz imaging in areas such as security, medical diagnostics, and material characterization. 
Further innovation of CMOS-based THz imaging circuits will be essential for the proliferation of high-resolution sensors across applications \cite{byreddy2023lensless,zhu2022_426ghz}. 

Groundbreaking developments in mmWave and THz radar imaging systems highlight the ongoing efforts to harness the unique properties of these frequency ranges for meaningful imaging and sensing applications across a broad set of industries. 
The massive bandwidths and higher operating frequencies achieved by these systems enable improved spatial sensing capabilities, as evident in (\ref{eq:planar_spatial_resolution}) and (\ref{eq:cylindrical_spatial_resolution}), with smaller platforms. 
The development of novel transceivers and radar systems across various frequencies and bandwidths has contributed to the progress in mmWave and THz imaging, bringing us closer to the realization and proliferation of advanced imaging systems in both industrial and research contexts \cite{o2023perspective}. 



\section{Data-Driven SAR Algorithms}
\label{sec:ml_sar_algorithms}

In this section, we review ongoing innovations and highlight the challenges for emerging data-driven SAR algorithms across several applications. 
SAR data are inherently distinct compared with optical images on several fronts, and exploiting the nature of these signals is crucial to enable robust and generalizable machine learning and deep learning approaches to near-field SAR imaging problems. 

\subsection{Introduction to Learning Techniques for Applications in High-Resolution Radar Perception}
\label{subsec:ml_radar}

Data-driven algorithms have been applied to a variety radar sensing tasks over the past decade, starting spanning human-computer interaction (HCI), something, and something else. 

mmWave radar has been employed in conjunction with neural network models to classify hand motions \cite{kim2016hand,suh2018_24GHz_gesture}, hand poses \cite{kim2017staticgesture,smith2021sterile}, perform hand tracking \cite{smith2021An}. 
In \cite{kim2016hand}, a CNN technique was applied to radar data to classify dynamic, moving hand gestures, but was not robust when new users were introduced. 
The ability to generalize to a diverse set of users is essential for a detection and classification algorithm in real-world applications. 
To overcome this issue, a real-time gesture classification algorithm was developed using a 24 GHz FMCW radar and a long short-term memory (LSTM) network \cite{suh2018_24GHz_gesture}, a variant of the RNN. 
However, the issue of generalizability remains a central issue for hand gesture detection and recognition. 
In \cite{leem2020detecting}, digit-writing technique is introduced that yields improved generalizability based on a CNN model. 
Later, \cite{zheng2021dynamic} introduced a transformer-based architecture for hand gesture classification in vehicle environments on a 60 GHz radar platform. 
The transformer-based approach yielded superior numerical performance compared with CNN and CNN-LSTM models. 
The preprocessing applied to the radar signals allows for multiple dimensions of spatial feature extraction. 
Using a common CNN, several preprocessing techniques are investigated and their impact on model robustness and computational complexity \cite{smith2022novel}.  
Static hand gesture recognition, a more challenging problem compared with the dynamic case as the classes tend to be less distinct, has been explored in the literature. 
Support vector machine (SVM) classifiers were compared to CNN models in \cite{kim2017staticgesture}, demonstrating the superiority of a wider CNN algorithm compared with a traditional SVM. 
In \cite{smith2021sterile}, a sterile training method was proposed that improves classification accuracy by employing ``sterile'' data collected from highly reflective cutouts in the shape of each pose that emphasize the unique features of each class. 

Data-driven radar perception is a promising solution to privacy-centric human activity monitoring. 
Human activity recognition is a valuable application for several fields \cite{ramasamy2018recent}. 
By detecting and monitoring movement, this technology allows for improved healthcare, detection of falls or irregular behavior, and reducing response times for emergency assistance by leveraging Doppler and micro-Doppler signatures \cite{khan2016ram}. 
A recent study \cite{abedi2022non} applies mmWave radar for wellness monitoring in long-term care facilities. 
In \cite{abedi2023deep} a vehicle in-cabin safety monitoring system is proposed using deep learning to detect, count, and classify passengers. 
The proposed technique is capable of rapidly identifying unattended infant and children passengers, identifying occupied seats, and classifying passengers to improve in-vehicle safety. 

Similarly, deep learning-assisted radar imaging poses several advantages, including privacy and depth sensing, compared to optical cameras for human body tracking. 
Contactless human pose estimation is a vital technology for electronic human expression, particularly with the rise of AR/VR applications. 
In \cite{sengupta2020mm}, a human posture detection algorithm was introduced based on a CNN-model for real-time estimation. 
State-of-the-art 3-D single subject pose tracking has since been challenged by \cite{lee2023hupr} and multi-subject pose estimation has been investigated showing promising results \cite{kong2022m3track}. 
Additionally, radar has shown promise for contactless and camera-less facial reconstruction. 
A recent study \cite{xie2023mm3dface} proposed a facial expression tracking algorithm using a single mmWave radar capable of recovering facial orientation, position, and subtle expressions. 
However, as THz systems become increasingly affordable and commercially viable, the additional sensors and system bandwidth could significantly improve existing solutions. 

Furthermore, learning techniques have been successful for both mmWave and THz concealed threat detection. 
Passive mmWave \cite{lopez2017deep,lopez2018using} and THz \cite{xu2022YOLOMSFG} threat detection algorithms have been developed to achieve real-time detection of concealed objects using segmentation \cite{lopez2017deep} and you only look once (YOLO) multi-scale filtering and geometric (MSFG) augmentation techniques \cite{xu2022YOLOMSFG}. 
Compared to passive imaging, which typically uses large real array radar (RAR) \cite{lopez2018using}, conventional concealed object detection uses active imaging SAR to reduce costs. 
Active THz \cite{yuan2018a,xiao2018r} and mmWave \cite{yanik2018millimeter,smith2022ffh_vit,liu2019concealed,wei2021_3DRIED} systems are essential for rapid and effective concealed weapon detection. 
Whereas data-driven techniques for EM threat detection have commonly relied on conventional optical computer vision techniques, there are key differences in the characteristics of mmWave and THz signals and images compared with optical images that must be taken into account. 
Although some insights gained from optical computer vision can be extended to the RF domain, a thorough understanding of signal characteristics, imaging techniques, and complex data representation is essential for the success and adoption of future array imaging technology. 

\subsection{Complex-Valued Data Representation in Data-Driven Radar Signal Processing Algorithms}
\label{subsec:complex_valued_data}

A fundamental issue when applying data-driven algorithms to EM data is the challenge of data representation. 
Specifically, EM signals contain useful information in both the magnitude and phase of the measurement. 
This can be observed also in the delicate handling of the phase during image reconstruction, as detailed in Section \ref{subsec:sar_imaging_algorithms}. 
Representing the inherently complex-valued data can be done in a variety of ways. 
However, implications of each representation technique must be considered when developing an algorithmic solution, particularly for data-driven applications.

The simplest solution to represent complex-valued data is to take the magnitude information only and discard the phase. 
Although some research has shown comparable performance using magnitude-only processing of radar data \cite{scarnati2021complex,vasileiou2022efficient,smith2022ffh_vit}, most applications exhibit superior performance using complex-valued neural network processing \cite{gao2018enhanced,jing2022enhanced,smith2021An}. 

A naive method for representing the complex-valued data is to layer the real and imaginary part as a two-channel representation \cite{smith2021An}. 
For example, suppose $\bm{z} = \bm{z}_R + j\bm{z}_I$ where $\bm{z}_R$ and $\bm{z}_I$ are real-valued. 
Using the layering technique, $\bm{z}_R$ and $\bm{z}_I$ are treated as independent channels of the input signal. 
This treatment retains some of the phase information at the input but renders its meaning ambiguous at intermediate layers of the network, which tend to have a larger number of feature channels. 
Although this seems like an elegant solution to leverage the conventional real-valued deep learning software, subsequent processing may not have the expected properties, particularly on phase-sensitive data like radar samples. 

\subsubsection{Complex-Valued Convolutional Neural Networks}
\label{subsubsec:complex_valued_convolutional_neural_networks_cv-cnns}
Rather than separating the real and imaginary parts, extensive work has been made towards fully complex-valued neural networks (CV-NNs) \cite{barrachina2023theory,lee2022complex,bassey2021survey}. 
However, the first CV-NN was proposed long before the modern advancements in deep learning architectures. 
In 1990, Kim and Guest introduced a complex-valued multi-layer perceptron (MLP) \cite{kim1990modification}. 
Later, Haensch and Hellwich extended this concept to incorporate the popular convolution operation by proposing a complex-valued convolutional neural network (CV-CNN) for object classification in polarimetric SAR (PolSAR) images \cite{haensch2010complex}. 
Since then, the popularity of CV-CNNs has increased substatially. 

However, implementing fully complex-valued convolution on modern hardware comes at a cost. 
Since the convolution operator is only implemented for real-valued data on current platforms, realizing complex-valued convolution requires four real-valued convolutions as follows. 
Assuming a complex-valued weight matrix $\bm{W} = \bm{W}_R + j \bm{W}_I$ is composed of two real-valued matrices- the real part $\bm{W}_R$ and imaginary part $\bm{W}_I$, the complex-valued convolution between $\bm{W}$ and $\bm{z}$ can be defined as
\begin{multline}
    \label{eq:complex_convolution}
    \bm{W} \ast \bm{z} = (\bm{W}_R + j\bm{W}_I) \ast (\bm{z}_R + j\bm{z}_I) = \bm{W}_R \circledast \bm{z}_R - \bm{W}_I \circledast \bm{z}_I + j(\bm{W}_R \circledast \bm{z}_I + \bm{W}_I \circledast \bm{z}_R),
\end{multline}
where $\ast$ is the complex-valued convolution operator and $\circledast$ is the real-valued convolution operator \cite{trabelsi2018deep}. 
Hence the real-valued convolution between the real and imaginary parts of both the weight matrix and input tensor must be computed, resulting in a total of four real-valued convolutions. 
Alternatively, Gauss' trick for complex numbers can be employed to reduce the computation to three real-valued convolutions \cite{chakraborty2019sur_real,chakraborty2022SurReal,smith2023complex}. 
However, this model retains the structure of the complex-valued data and dependency between magnitude and phase and has proven superior to a real-valued CNN (RV-CNN) for many applications \cite{gao2018enhanced,jing2022enhanced,wang2021rmistnet,sharma2023super}. 
Similarly, complex-valued transformer \cite{yang2020complex,dong2021signal} and attention \cite{kothapally2022complex} architectures have been introduced to perform the complex-valued counterpart of modern computer vision and natural language processing techniques. 

However, complex-valued neural networks require several additional considerations for implementation. 
Complex-valued valued backpropagation requires the computation of many gradients, or partial derivatives. 
However, derivatives can only be computed on analytical, or holomorphic, functions. 
Since Liouville's theorem states that CV-NNs will use non-holomorphic functions, complex-valued backpropagation requires Wirtinger calculus, which treats holomorphic functions only as a special case, to compute gradients  \cite{barrachina2023theory,trabelsi2018deep}. 
Fortunately, the industry standard, PyTorch, automatically handles the complex-valued backpropagation and Wirtinger calculus computation of the gradients internally. 

Activation functions must also be considered for complex-valued data. 
The most common complex-valued activation functions apply a traditional activation function to the real and imaginary parts of the data separately, as 
\begin{equation}
    \label{eq:general_split_activation_function}
    F(\bm{z}) = G_R(\bm{z}_R) + jG_I(\bm{z}_I),
\end{equation}
where $G_R(\cdot)$ and $G_I(\cdot)$ are nonlinear activation functions and typically $G_R(\cdot) = G_I(\cdot) = G(\cdot)$. 
For example $\mathbb{C}$ReLU$(\cdot)$ uses the rectified linear unit (ReLU) $G(\cdot) = \text{ReLU}(\cdot)$ \cite{gao2018enhanced}, $\mathbb{C}$PReLU$(\cdot)$ uses the parametric ReLU (PReLU) $G(\cdot) = \text{PReLU}(\cdot)$ \cite{jing2022enhanced,smith2023dual_radar}, and $\mathbb{C}$Tanh$(\cdot)$ uses the hyperbolic tangent $G(\cdot) = \tanh(\cdot)$ \cite{lee2022complex}. 
Other activation functions, such as the zReLU \cite{guberman2016zrelu} and modReLU \cite{arjovsky2016modrelu}, restrict activation to certain regions of the complex plane. 
Another popular complex-valued activation function the cardioid \cite{virtue2017cardioid} function, function is an extension of the ReLU that dampens the amplitude response based on the phase while retaining the phase information. 

Similarly, computing the loss on complex-valued data can be done simply by computing a standard metric, like L2-norm or SSIM \cite{wang2021tpssiNet}, on the real and imaginary parts separately. 
However, complex-valued specific metrics have been developed, such as Complex Cauchy error function and complex log-cosh error function \cite{prashanth2002surface}. 
Recently, Terpstra \textit{et al.} introduced a symmetric loss function for complex-valued data based on a geometrical interpretation on the complex plane, demonstrating promising results for magnetic response imaging (MRI) applications \cite{terpstra2022loss}. 

\subsubsection{Geometrical Approach to Complex-Valued Convolution and the Weighted Frech\`et Mean}
On the other hand, recent progress has been made towards geometrically derived deep learning models specifically designed for complex-valued data. 
Specifically, the convolution operator has gained immense popularity following AlexNet \cite{krizhevsky2012imagenet} due to the spatial equivariance property, allowing the convolutional layer to produce spatially translated outputs based on a spatially translated inputs. 
In \cite{chakraborty2019sur_real}, a novel convolution operator was introduced for complex-valued data that demonstrates equivariance to the product group of planar rotation and non-zero scaling. 
By approaching the manifold of complex numbers, $\mathbb{C}$, as the product space of the positive real numbers and the rotation group $SO(2)$, the weighted Frech\`et mean (wFM) is introduced to replace the standard Euclidean convolution operator. 
The proposed convolution operator demonstrates equivariance to complex scaling, non-zero magnitude scaling and rotation (phase shift), unlike the traditional Euclidean convolution operator in (\ref{eq:complex_convolution}). 
Whereas for conventional convolution, the values of weight matrix $\bm{W}$ can be arbitrary, the values of the wFM weight matrix must be non-negative and sum to 1. 
For complex-valued data, the wFM operator reduces to standard convolution with a convexity constraint on the values of the weights \cite{chakraborty2022SurReal}. 

The wFM complex-valued convolution operator shows significant promise in reducing model size for CV-NNs. 
It has already been employed in several techniques yielding superior numerical performance to their Euclidean convolution counterparts with a fraction of the model parameters. 
the wFM-based CNN is applied to remote sensing SAR image classification \cite{keydel1996mstar}, RF modulation classification \cite{oshea2016radioml}, and RF fingerprinting \cite{brown2021charrnets} demonstrating impressive performance with significantly smaller model sizes \cite{chakraborty2019sur_real,chakraborty2022SurReal}. 
As noted in \cite{scarnati2021complex}, however, the additional computation of the weighted Frech\`et mean increases the networks' computational burden during training. 

Despite the notable strengths of the wFM as a convolution operator, it has not gained significant footing for SAR image processing, particularly for near-field image processing and image super-resolution. 
Compared with Euclidean convolution as in (\ref{eq:complex_convolution}), wFM-based CV-CNNs are yet to be widely adopted for data-driven RF signal processing. 
This remains a promising future research direction for THz and mmWave array processing and has the potential for widespread impact on the broader data-driven signal processing community if explored further. 

\begin{figure}[ht]
    \centering
    \includegraphics[width=0.55\textwidth]{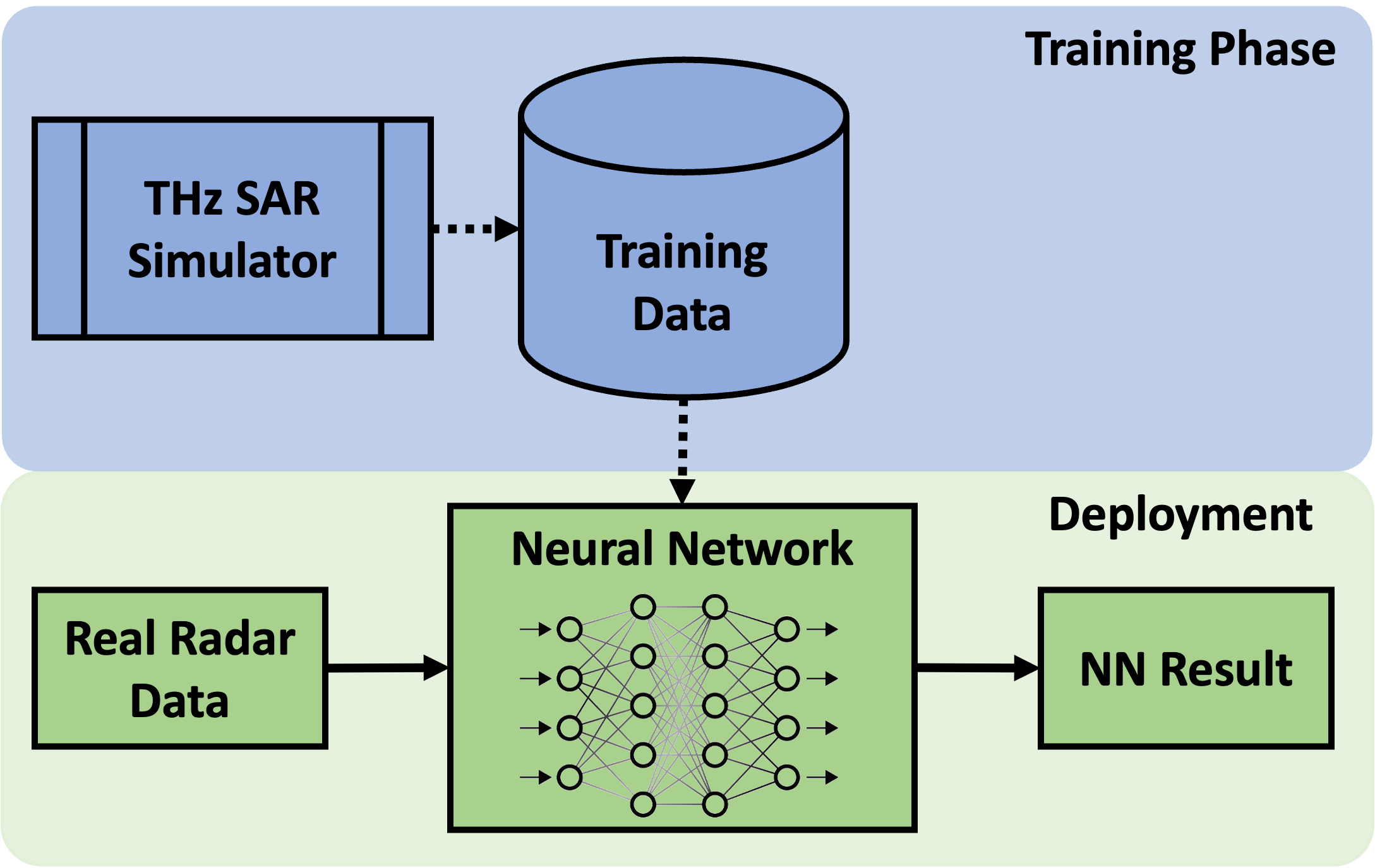}
    \caption{Overview of the common training regime for deep learning-based radar image processing algorithms. Generating meaning SAR data for training is essential for developing algorithms that generalize well to real-world data. The neural network is first trained using synthetic training data produced using a simulator such as the THz SAR simulator proposed in Section \ref{sec:thz_sim}. The model can be tested on simulated and/or real data before being deployed on real radar data.}
    \label{fig:synthetic_training_pipeline}
\end{figure}

\subsection{Training Deep Learning Models for SAR Image Processing}
\label{subsec:training_deep_learning_models_for_SAR_image_processing}

Whereas deep learning on SAR images has seen tremendous success for remote sensing and satellite SAR applications \cite{li2023survey,zhu2021deep,haensch2010complex,keydel1996mstar}, it is still maturing for mmWave and THz SAR image processing. 
In addition, as with other traditional computer vision tasks \cite{li2021iGibson}, because collecting and labeling near-field SAR images is challenging, mmWave and THz image super-resolution relies heavily on synthetic data.
The typical pipeline employed for deep learning SAR image processing algorithms is detailed in Fig. \ref{fig:synthetic_training_pipeline}. 
The model is trained using simulated data, testing using either simulated or real data, and deployed to operate on real data \cite{gao2018enhanced,smith2022ffh_vit,zhang2019target,li2020adaptive}. 
For this sim-to-real regime, the generalizability of the networks to real SAR data depends largely on the quality of the synthetic data. 
Due to lack of existing benchmark datasets and widely available tools, prior research relies on the authors' ability to independently develop meaningful datasets. 
Although the NN result could be a set of classification probabilities, super resolved image, or any other metric meaningful for a downstream task, the burden currently falls on researchers to create their own datasets. 
This is a tremendous opportunity for the development of mmWave and THz tools for realistic image synthesis. 
By reducing need to develop custom data for each research effort, greater attention can be paid to forging novel algorithms and developing explainable data-driven algorithms for radar image processing. 

\begin{figure}[ht]
    \centering
    \includegraphics[width=\textwidth]{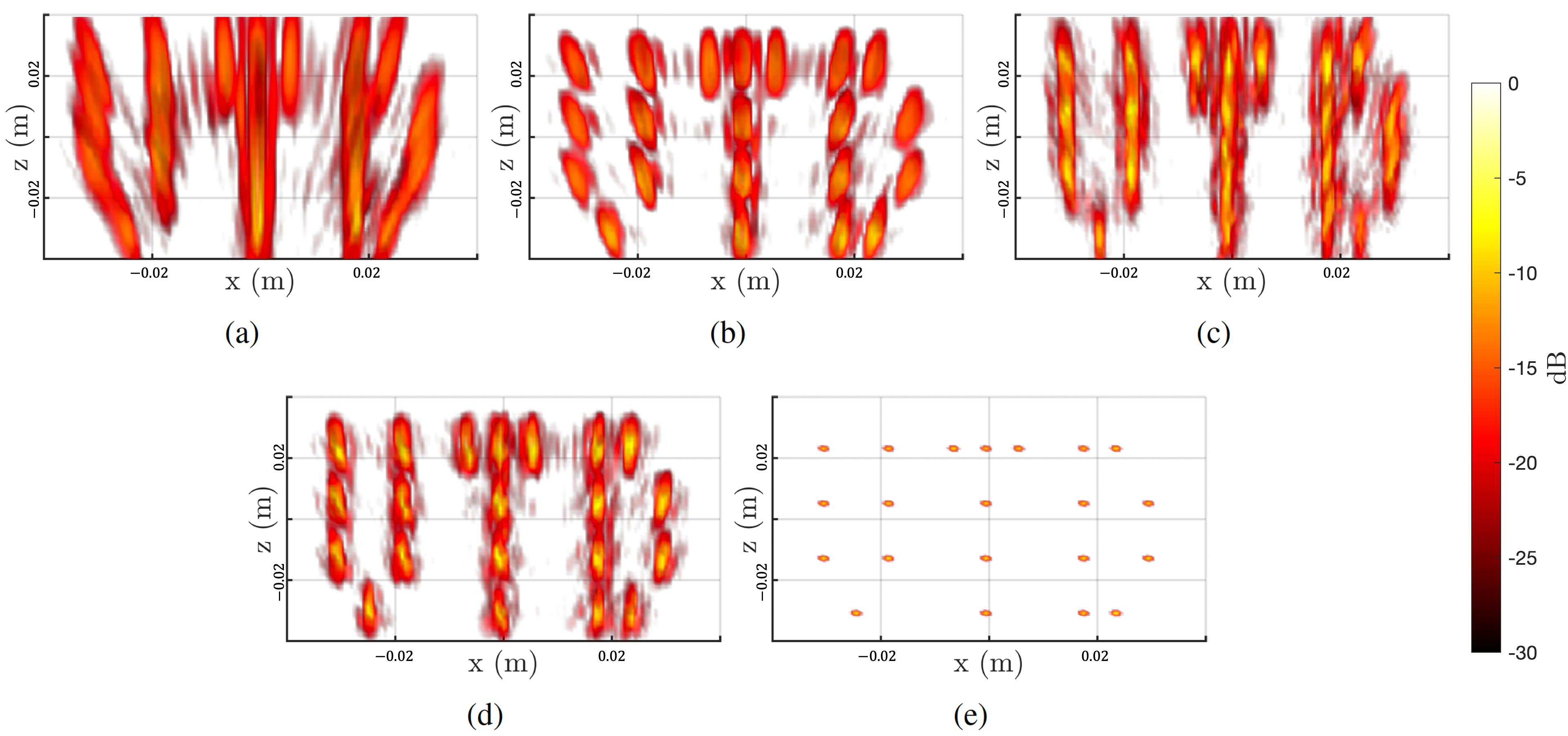}
    \caption{Simulation imaging results of ``UTD'' shape with varying system parameters using THz toolbox illustrating the spatially variant distortion and impact of bandwidth and aperture size on image resolution. Using THz transceiver with center frequency of 435 GHz and linear synthetic aperture with spacing of $\lambda/4$, as shown in Fig. \ref{fig:rectilinear_system_geometries}a. Configurations: (a) 128 element array with 5 GHz bandwidth, (b) 128 element array with 10 GHz bandwidth, (c) 256 element array with 5 GHz bandwidth, (d) 256 element array with 10 GHz bandwidth, and (e) ground-truth ideal image generated with prior knowledge of the target.}
    \label{fig:utd_res} 
\end{figure} 

In particular, the issue of synthetic SAR image content remains widely oversimplified in existing research.  
Typically, training data for mmWave and THz SAR are generated by assuming that the scene consists of randomly placed point scatterers \cite{gao2018enhanced,fan2021fast,smith2021An,sharma2023super}; however, the image quality degrades in applications when the targets are not point-like and the shape of the target is important, for example, concealed threat detection, packaging, etc. \cite{yanik2020development,smith2023dual_radar,liu2019concealed}. 
The toolbox presented later in this article provides an efficient solution for generating large synthetic datasets with diverse targets by allowing the user to import 3-D objects, see Fig. \ref{fig:cylindrical_example}, and improves accessibility for researchers new to THz SAR principles. 

\subsection{Deep Learning Single-Image SAR Super-Resolution Techniques}
\label{subsec:sar_super_resolution}

A task of particular interest to the mmWave and THz imaging communities is near-field SAR super-resolution. 
Following (\ref{eq:planar_spatial_resolution}) and (\ref{eq:cylindrical_spatial_resolution}), the spatial resolution of near-field SAR is determined by the extent of the synthetic aperture, the operating frequency, and the system bandwidth. 
As discussed in Section \ref{sec:sar_algorithms}, for many high-resolution imaging tasks, the limitations on spatial resolution imposed by physical constraints, such as aperture size and bandwidth-limited hardware, are a fundamental issue for recovering high-fidelity 3-D images \cite{dai2021imaging,jing2022enhanced}. 
The primary challenge in SAR super-resolution is to overcome the artifacts caused by system limitations in the form of distortion, blur, noise, and aliasing.  

SAR super-resolution algorithms were first introduced for far-field image enhancement \cite{gao2018enhanced}. 
However, typical far-field SAR signal models assume a spatially invariant distortion model and a small number of simplistic point target reflectors, practically modeling the problem as a spectral super-resolution problem, as addressed in  \cite{izacard2019aLearning,pan2021complexFrequencyEstimation}. 

For near-field mmWave and THz imaging applications, artifacts due to system limitations are highly nonlinear and spatially variant.
Fig. \ref{fig:utd_res} demonstrates this phenomenon by comparing reconstructed images of the ``UTD'' shape using different SAR configurations \cite{smith2022efficient}. 
Each permutation of the 5 GHz and 10 GHz bandwidths with the 128 element and 256 element linear arrays are shown to illustrate the impact of bandwidth and aperture size on image resolution and spatial variance of the distortion introduced by these system limitations. 
As evident in Fig. \ref{fig:utd_res} and according to the relations in (\ref{eq:planar_spatial_resolution}), larger bandwidth and increased aperture size result in finer range ($z$) and cross-range ($x$) resolution, respectively. 
To overcome these challenges inherent to near-field SAR imaging, super-resolution and image restoration models must be properly designed to leverage the unique characteristics of EM images. 
Algorithms for near-field SAR super-resolution have been introduced using CNN \cite{huang2019through,jing2022enhanced} and generative adversarial network techniques \cite{zhang2019target,vasileiou2022efficient} for mmWave imaging. 
The conventional signal processing approach for super-resolution assumes sparsity constraints in a certain domain and uses a compressed-sensing (CS) algorithm to recover the image. 
Recent research compares data-driven SAR super-resolution to conventional CS techniques advantageously \cite{cheng2020compressive,gao2018enhanced}. 
Furthermore, hybrid-techniques have been introduced using deep learning to generate a sparsifying basis for a CS algorithm, which drastically reduces computational complexity. 
Such approaches employ fast iterative shrinkage-thresholding algorithm (FISTA) \cite{wang2020csrnet} and approximate message passing (AMP) \cite{wang2021tpssiNet} algorithms in conjunction with data-driven techniques to achieve superior performance than conventional algorithms, such as orthogonal matching pursuit (OMP) and RMA. 
Such hybrid-learning approaches offer a novel method for solving imaging problems by exploiting the nature of the signals in conjunction with a data-driven algorithm.
Hence, supposed physical limitations are being surpassed as machine learning and deep learning algorithms are proving capable of learning the innate characteristics of mmWave and THz signals.

Similarly, super-resolution algorithms have been developed that operate on images recovered by the RMA or similar Fourier-based techniques (PSM, CSA, etc.) to improve resolution. 
In \cite{dai2021imaging}, enhancement algorithm for mitigating multistatic MIMO artifacts on near-field images was introduced.  
Jing \textit{et al.} proposed a complex-valued CNN (CV-CNN) architecture to achieve state-of-the-art SAR super-resolution on mmWave images \cite{jing2022enhanced}. 
In \cite{smith2022ffh_vit}, a mobile vision transformer \cite{mehta2021mobilevit} architecture was introduced to overcome the distortion due to position estimation errors in near-field SAR systems. 

THz SAR imaging has witnessed a similar trend towards deep learning-based image super-resolution with an emphasis on image domain enhancement algorithms. 
In \cite{han2020terahertz}, an unsupervised ``zero-shot'' learning technique was introduced, where super-resolution is performed using a network trained only on the input image itself, showing robustness for complicated solid targets. 
A supervised technique, trained using synthetic data, was introduced in \cite{li2020adaptive} for 2-D THz SAR enhancement that outperforms conventional super-resolution techniques, such as the Lucy-Richardson algorithm and normalized sparsity measurement (NSM) blind-deconvolution. 
An algorithm for 3-D THz SAR super-resolution was proposed using a CNN framework in \cite{fan2021fast}; however, since the model was trained on simulated THz images consisting of exclusive point targets, the recovered images from solid objects tend to lose fidelity and appear point-like rather than solid physical objects. 

\begin{figure}[ht]
    \centering
    \includegraphics[width=0.65\textwidth]{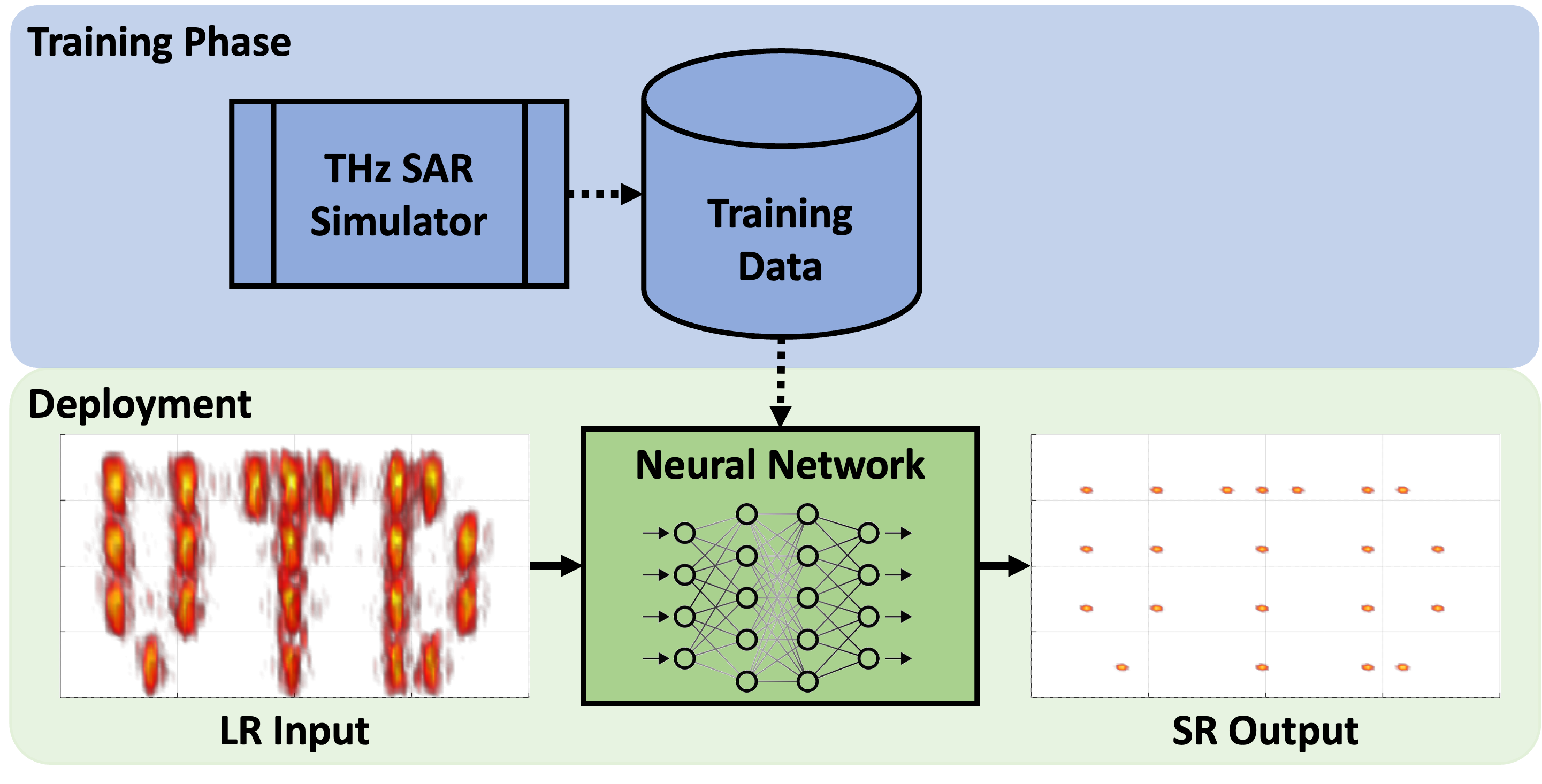}
    \caption{Training regime for SAR image super-resolution using synthetic data for training.}
    \label{fig:synthetic_superresolution_training_pipeline}
\end{figure}

As shown in Fig. \ref{fig:synthetic_superresolution_training_pipeline}, the typical training strategy for SAR image super-resolution is a special case of the generalized pipeline detailed in Fig. \ref{fig:synthetic_training_pipeline}. 
A simulated training dataset is synthesized comprising low-resolution (LR) input features and high-resolution (HR) corresponding labels. 
Using the proposed toolbox, LR and ground-truth HR images can be easily simulated using point-like scatterers or solid 2-D or 3-D shapes, enabling rapid generation of realistic large SAR image datasets. 
In deployment, the trained neural network receives real LR images and performs the learned super-resolution to produce super-resolved (SR) output images. 
However, as discussed in Section \ref{subsubsec:rma}, image data before, after, and throughout the image reconstruction may be leveraged by signal processing and deep learning algorithms for improving network robustness.  

\begin{figure}[ht]
    \centering
    \includegraphics[width=0.65\textwidth]{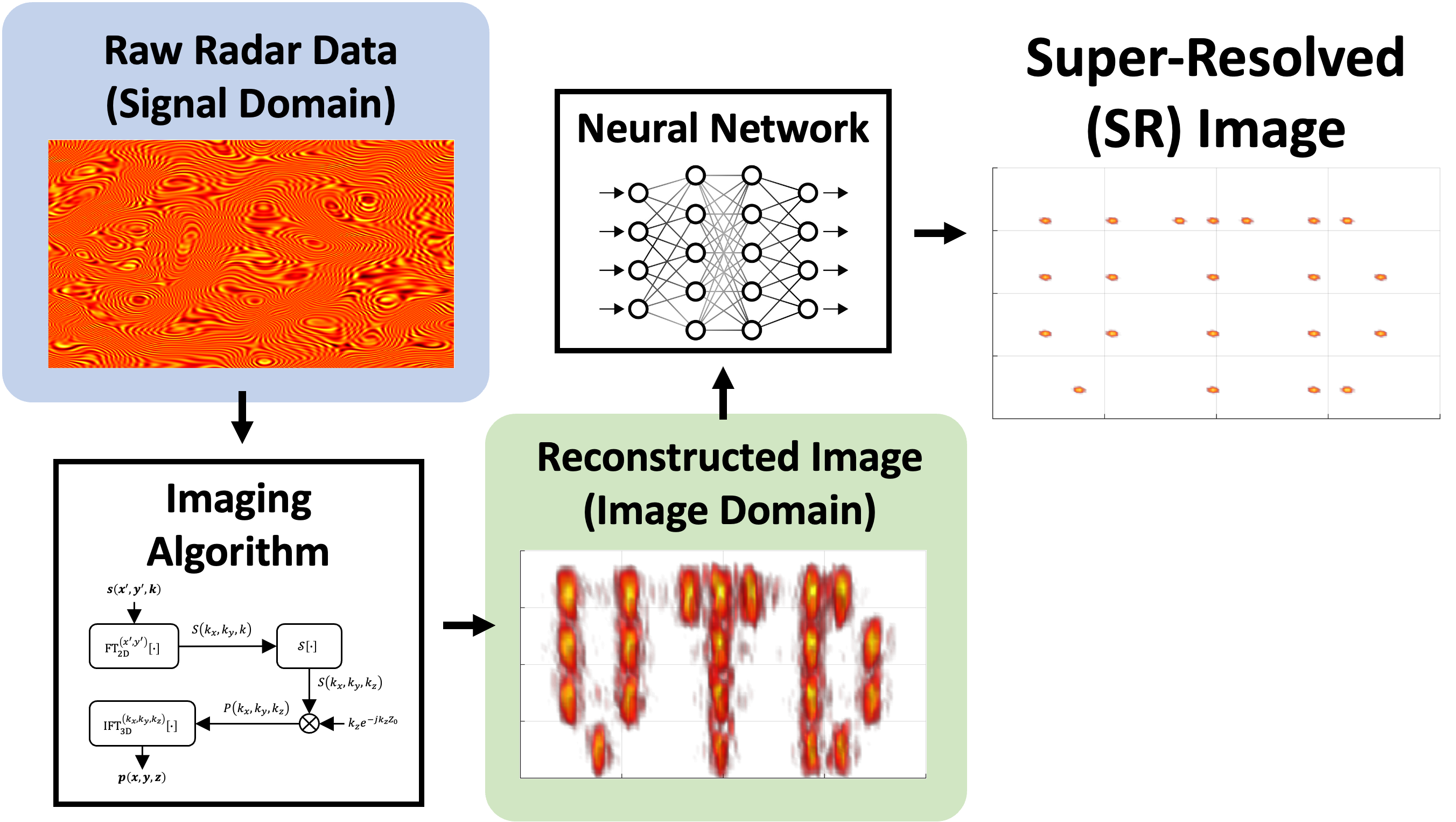}
    \caption{Signal domain data, shown in blue, consists of the data prior to the image reconstruction algorithm. Image domain data are the data after the image is recovered. Intermediate domain data exist at various intermediate steps in the reconstruction process between signal processing operations.}
    \label{fig:signal_domain_vs_image_domain}
\end{figure}

\subsection{Signal, Image, and Intermediate Domain Processing}
\label{subsec:signal_domain_vs_image_domain}

The conventional pipeline for applying data-driven techniques to SAR images, employed in \cite{jing2022enhanced,gao2018enhanced,sharma2023super,smith2021An,smith2022ffh_vit,vasileiou2022efficient,li2020adaptive,fan2021fast,dai2021imaging} and shown in Fig. \ref{fig:synthetic_superresolution_training_pipeline}, operates on the images recovered using image reconstruction algorithms such as the RMA. 
Alternatively, some recent efforts have proposed algorithms applied directly to the raw radar data with minimal preprocessing. 
We propose defining the raw radar data prior to the image reconstruction algorithm as \textit{signal domain} data, and images recovered by the focusing algorithms as \textit{image domain} data, as shown in Fig. \ref{fig:signal_domain_vs_image_domain}. 
Additionally, data throughout various steps of the reconstruction algorithm are referred to as \textit{intermediate domain} data. 

In \cite{wang2018wavenumber}, a signal processing technique is proposed leveraging the matrix pencil algorithm to fuse signals from multiple radars at the signal level. 
However, the proposed method is incapable of reconstructing signals consisting of solid targets and is insufficient for many applications. 
A deep learning variant was recently introduced capable of fusing multiband signals for complex target scenes with great success on real and simulated data \cite{smith2023dual_radar}. 
The proposed technique relies on a spectral understanding of multiband signal fusion and the sinc-effect and operates in both the sample domain and its spectral dual. 
The algorithm detailed in \cite{smith2023dual_radar} is a signal domain algorithm since it operates exclusively on the raw radar samples and images are reconstructed using the processed data. 
On the other hand, the techniques introduced in \cite{sharma2023super,fan2021fast,jing2022enhanced,li2020adaptive,dai2021imaging} are a more common image domain algorithms as they operate on data after image reconstruction. 

However, at the time of this publication, end-to-end models that operate in the signal domain, intermediate domains, and image domain remain unexplored in existing literature. 
An end-to-end imaging algorithm operating on across domains could address issues such as finite aperture affects, clutter, device nonlinearities, and clock synchronization issues, while exceeding the theoretical bounds of spatial resolution \cite{smith2021An}. 
Recent work has proposed similar hybrid techniques that leverage the advantages of conventional signal processing methods in conjunction with emerging data-driven algorithms to improve the sensing pipeline \cite{pan2021complexFrequencyEstimation,wang2021tpssiNet}; however, this future direction has much unexplored potential. 
Using an interleaved hybrid technique to leverage both signal processing and data-driven algorithms may yield significant performance gains compared to current state-of-the-art algorithms while offering computational or numerical advantages, as demonstrated by \cite{pan2021complexFrequencyEstimation,wang2020csrnet,wang2021tpssiNet}. 
However, there are several challenges to implementing hybrid or end-to-end algorithms, including computational complexity, data representation, generalization, and loss function definition. 
Domain choice will likely be application dependent as different image processing problems face unique challenges and leveraging information at each domain may yield varying performance improvements depending on the task. 




\subsection{Challenges and Opportunities for Data-Driven SAR Algorithms}
\label{subsec:sar_challenges}

Data-driven radar image processing algorithms pose a promising solution to highly nonlinear problems such as image classification, object detection, super-resolution, denoising, and restoration but face several challenges to being widely adopted for commercial applications. 
These opportunities, in addition to those discussed throughout this section, are promising routes for further innovation in deep learning signal processing. 

Of the algorithms discussed, a strong emphasis towards CNN frameworks is noted. 
However, recent advancements in the computer vision community have seen the influence of transformer architecture, originally developed in the NLP community, on vision tasks \cite{dosovitskiy2020image_ViT,mehta2021mobilevit}. 
The swin transformer \cite{liu2021swin} demonstrated state-of-the-art numerical performance for several high-level tasks, while boasting reduced computational complexity compared with existing CNN-based methods. 
Transformer architectures have recently been explored in the literature for hand gesture recognition \cite{zheng2021dynamic} and object detection \cite{bai2021radar}, but they pose a significant opportunity for improving SAR imaging performance, which is yet to be investigated. 
Furthermore, application of wFM-based CV-CNNs, as discussed previously, has shown tremendous promise across several complex-valued RF problems yet remains unexplored for mmWave and THz SAR imaging. 
The adoption of modern computer vision and CV-NN methods is likely to stimulate acceleration of research results and industry adoption of data-driven signal processing algorithms.  

Additionally, whereas the computer vision community has established several meaningful benchmarks for accelerating research towards image classification \cite{krizhevsky2012imagenet}, segmentation \cite{lin2014microsoft}, super-resolution \cite{lim2017enhanced}, etc., formal benchmarks do not currently exist for standardizing the performance of mmWave and THz SAR imaging algorithms, making performance comparisons challenging. 
The introduction of common EM imaging benchmarks for various tasks is essential to the ongoing success and adoption of mmWave and THz imaging and is a promising route for future publications. 
The toolbox proposed in this article provides a solution for many researchers to quickly develop meaningful datasets for a variety of SAR image processing applications, but the proliferation of such tools is essential for rapid innovation in this field. 
Recently, datasets for passive \cite{lopez2018using} and active \cite{wei2021_3DRIED} mmWave imaging have been released, consisting of raw SAR data; however, without labels for these data, they cannot be used for supervised learning. 
To overcome these deficiencies, unsupervised learning methods \cite{han2020terahertz} may offer some solutions but must be more thoroughly investigated for EM signals.


\section{THz Imaging Toolbox with Interactive User Interface}
\label{sec:thz_sim}
In this section, we introduce a software platform and an interactive user interface for near-field mmWave and THz imaging. 
The proposed framework provides user control over the waveform signal, the EM characteristics of the antennas, array structure, SAR scanning geometry, target scene, and image reconstruction. 
Furthermore, it enables rapid dataset generation for deep learning tasks using sophisticated, intricate target scenes, and automatically generates ground-truth labels for each sample. 
The imaging toolbox includes an interactive user interface, API, and MATLAB source code and enables rapid system prototyping, reconstruction algorithm development, and training dataset construction. 
Our simulation platform and user interface have already proven useful to recent research efforts \cite{smith2021An,smith2020nearfieldisar,smith2022ffh_vit,smith2022efficient,vasileiou2022efficient,gezimati2022circular,gezimati2023curved,smith2022novel}, and may be impactful for a broad array of researchers and practitioners. 

\begin{figure}[ht]
    \centering
    \includegraphics[width=0.6\textwidth]{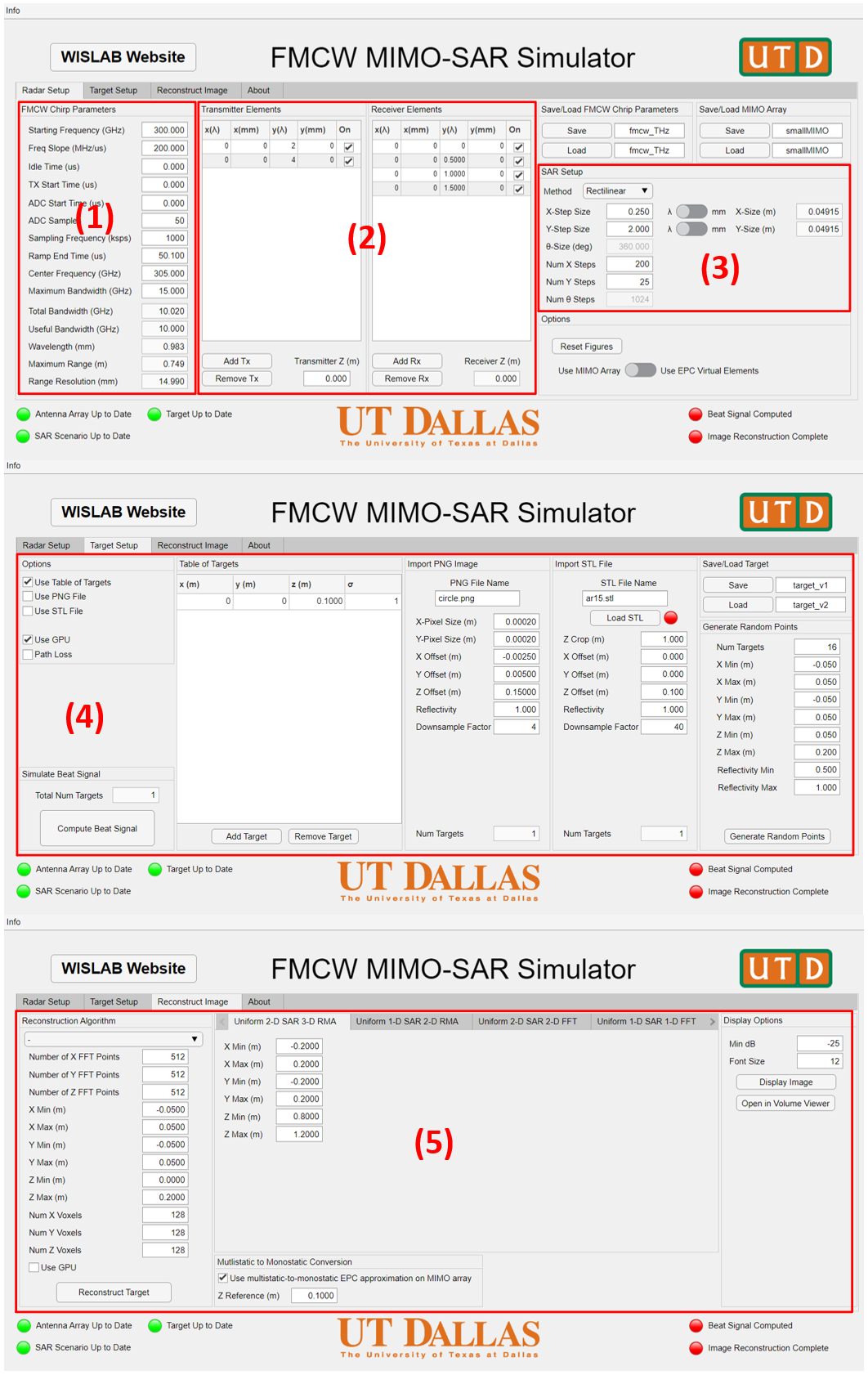}
    \caption{Five step process in the interactive GUI: (1) waveform properties, (2) antenna array configuration, (3) SAR scanning geometry, (4) target scenario, (5) image reconstruction.}
    \label{fig:gui_steps}
\end{figure}

\subsection{Operating the THz Imaging Toolbox}
\label{subsec:operating_thz_toolbox}
The operation steps of the proposed software user interface are illustrated in Fig. \ref{fig:gui_steps}. 
First, the user can configure the signal parameters to the desired frequency range, bandwidth, and sampling criteria for efficient waveform development. 
Next, the antenna elements and array geometry are fully customizable, allowing the user to determine the antenna gain pattern or import a custom gain pattern from MATLAB or an EM software such as HFSS, in addition to designing a custom array in any monostatic or multistatic geometry. 
Following Section \ref{subsec:sar_scanning_modes}, the proposed platform provides the four SAR scanning modes shown in Figs. \ref{fig:rectilinear_system_geometries} and \ref{fig:polar_system_geometries}, allowing the user to specify the sampling conditions and scanning dimensions. 
To promote diversity of images beyond simple point scatterers, any 3-D model can be imported for imaging simulation, in addition to other 2-D and point structures, allowing for meaningful datasets to be efficiently generated for machine learning purposes. 
Finally, the algorithms discussed in Section \ref{subsec:sar_imaging_algorithms} are implemented and made open-source to stimulate additional research and improve the accessibility of SAR concepts to the wider community. 
The proposed toolbox provides a interactive user interface in addition to an API allowing users across many experience levels to simulate meaningful THz and mmWave SAR scenarios. 

\subsection{Illustrative Example of THz Imaging Toolbox Simulation Capabilities}
\label{subsec:thz_toolbox_example}
\begin{figure}[ht]
\centering
    \begin{tabular}{c}
     \centering
     \includegraphics[width=0.45\textwidth]{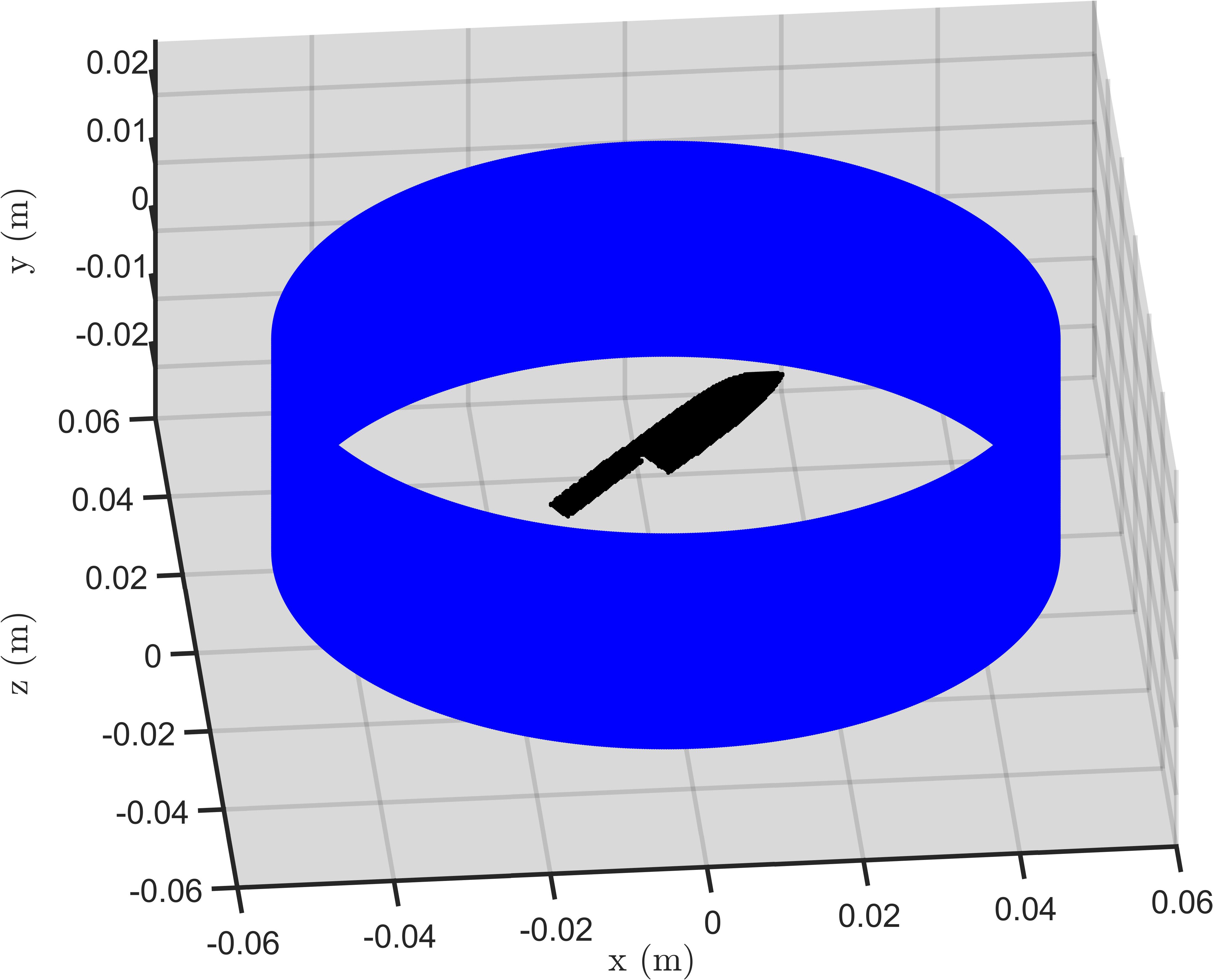} 
        \small(a)
    \end{tabular}
    \begin{tabular}{c}
     \centering
     \includegraphics[width=0.4\textwidth]{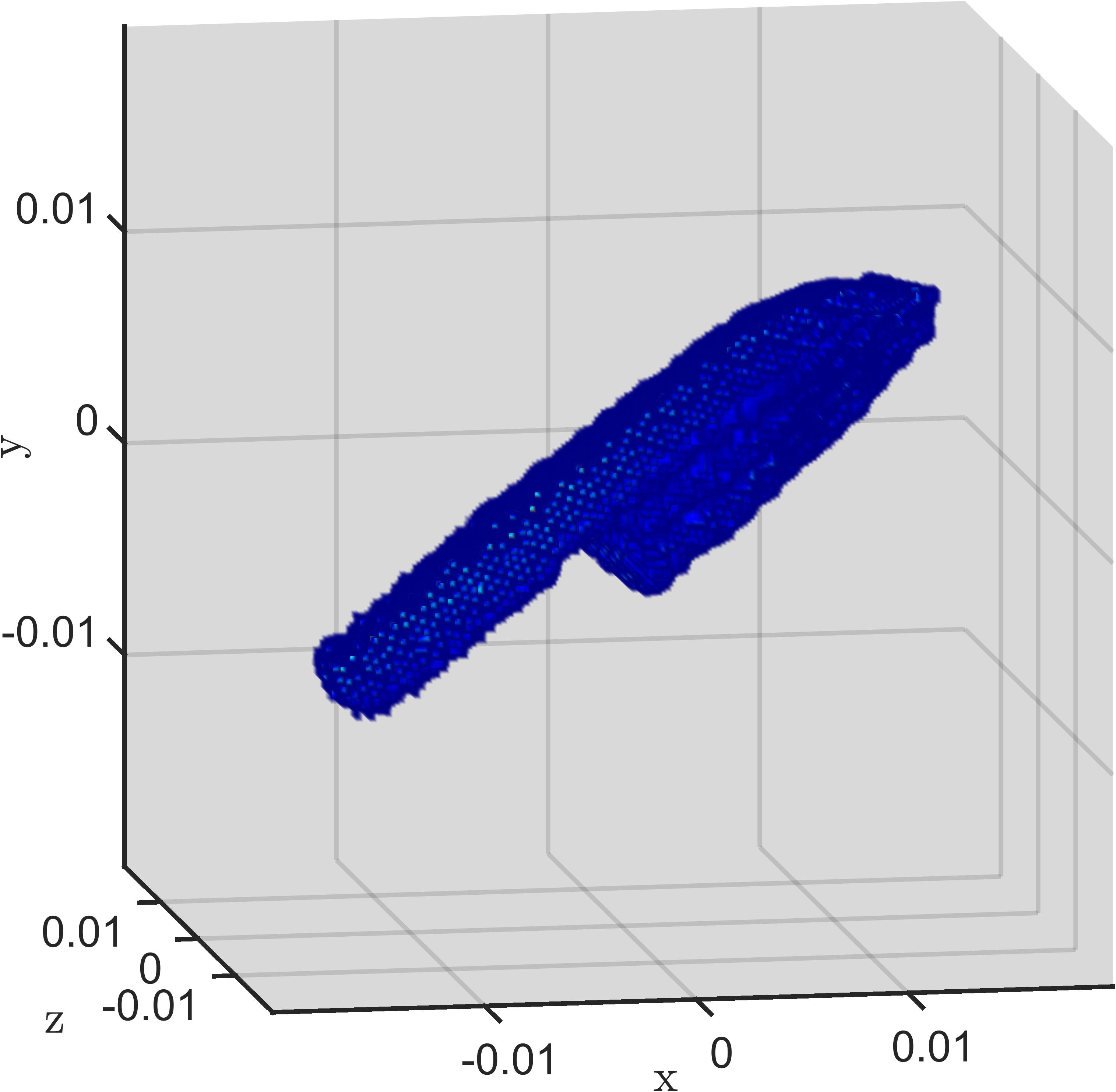} 
        \small(b)
    \end{tabular}
\caption{Example illustrating the software functionality for simulating an realistic 3-D model of a knife. (a) Scenario geometry of the cylindrical synthetic aperture and knife 3-D model. (b) Reconstructed image using the cylindrical RMA from the simulated data.}
\label{fig:cylindrical_example}
\end{figure}

To demonstrate the simulation capabilities of the platform, an example of a monostatic cylindrical SAR scenario with a 3-D model of a knife is shown in Fig. \ref{fig:cylindrical_example}a. 
The simulation assumes a wideband transceiver operating at a center frequency of 435 GHz with a bandwidth of 10 GHz using an antenna gain pattern designed in HFSS. 
Using a parallelized approach, the radar return signal can be efficiently computed even with a synthetic aperture of $128 \times 1024$ elements along the vertical and rotational directions, respectively. 
The 3-D image can be recovered using the cylindrical RMA algorithm, which is included in the toolbox and detailed in Section \ref{subsec:sar_imaging_algorithms}. 
As shown in Fig. \ref{fig:cylindrical_example}b, the recovered image is a high-fidelity 3-D reconstruction of the knife model; however, the spatial resolution is constrained by system limitations. 
The proposed software enables the simulation of realistic and intricate objects for mmWave and THz imaging scenarios. 
Furthermore, functionality is provided to automatically generate a ground-truth label for the 3-D model using a conventional model \cite{smith2021An,jing2022enhanced,dai2021imaging,gao2018enhanced}. 
Hence, large meaningful datasets can be rapidly developed for a host of data-driven tasks including image super-resolution, object detection, concealed item classification, and occlusion mitigation.

\subsection{Training Data-Driven Imaging Algorithms Using the THz Imaging Toolbox}
\begin{figure}[th]
    \centering
    \includegraphics[width=\textwidth]{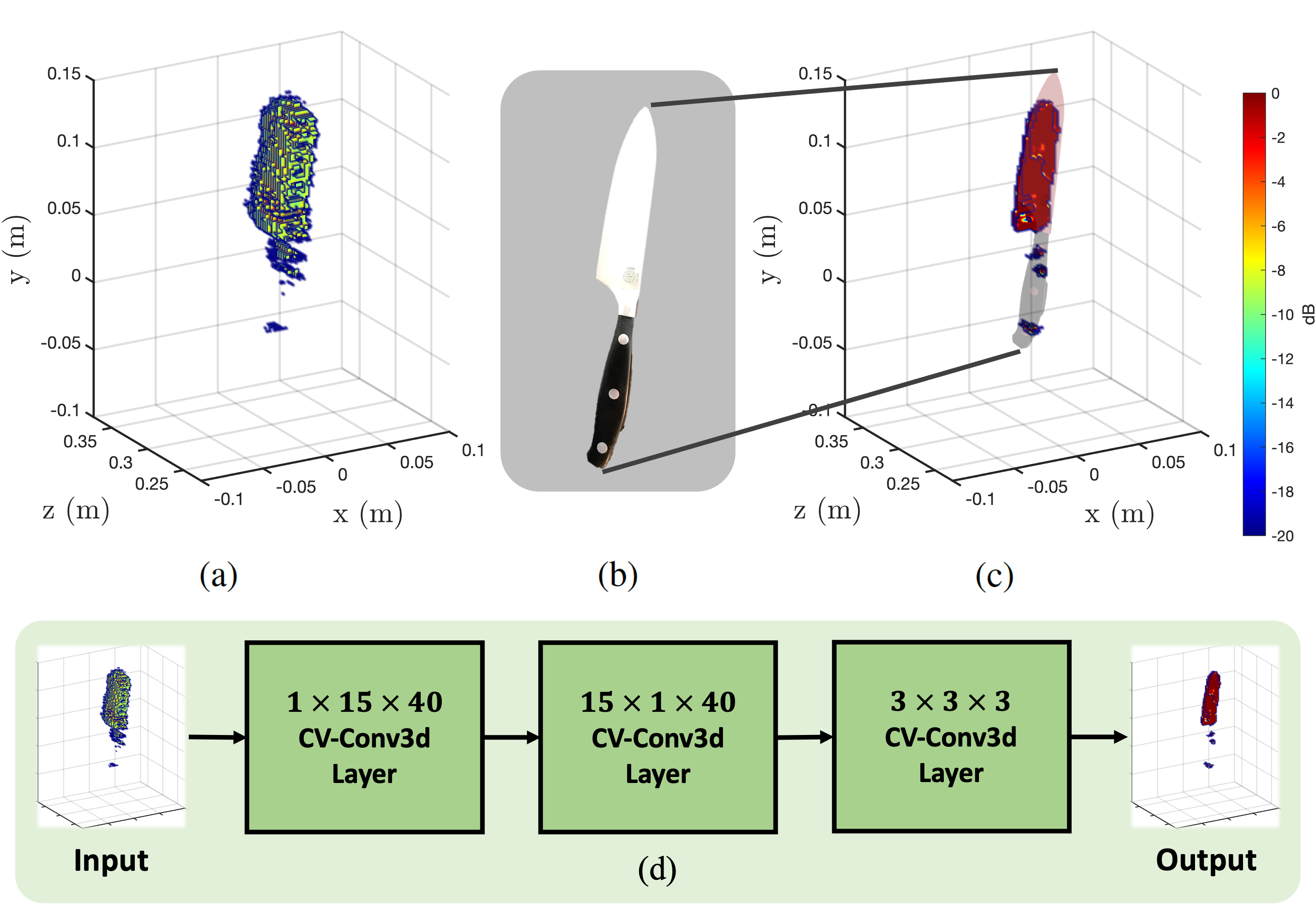}
    \caption{Example CV-CNN trained using data synthesized by the proposed THz imaging toolbox. (a) The image reconstructed using the planar RMA. (b) The kitchen knife being imaged by the planar scanner. (c) The super-resolved (SR) image after being processed by the trained network. (d) The CV-CNN architecture consisting of three 3-D CV-Conv layers taking the reconstructed RMA images as input and outputting estimated SR images.}
    \label{fig:knife_ex_cnn}
\end{figure}

The proposed THz imaging platform can be used to generate meaningful datasets for training mmWave and THz machine learning and deep learning algorithms. 
Earlier versions of the software were employed in \cite{smith2021An,smith2022ffh_vit,vasileiou2022efficient} to generate training datasets of 2-D SAR images for image enhancement in a variety of scenarios. 
Using the THz image toolbox, we create a dataset of 3-D SAR images with 20000 training and 3000 testing samples for the SAR super-resolution problem detailed in Fig. \ref{fig:synthetic_superresolution_training_pipeline}. 
Each SAR image is composed of solid targets, such as the knife shape shown in Fig. \ref{fig:cylindrical_example}, and randomly placed point targets. 
High-resolution ground-truth images, such as those shown in Fig. \ref{fig:utd_res}e, are also efficiently computed by the platform the simulation process, following the convention established in \cite{gao2018enhanced,vasileiou2022efficient,smith2022ffh_vit}. 
A simple CV-CNN architecture is proposed, as shown in Fig. \ref{fig:knife_ex_cnn}d. 
The network consists of three 3-D complex-valued convolution layers (CV-Conv) with varying kernel sizes. 
The first CV-Conv layer operates primarily along the horizontal cross-range ($x$) and range ($z$) directions with a filter size of $15 \times 1 \times 40$. 
Similarly, the kernel of the second CV-Conv layer is $1 \times 15 \times 40$ in order to operate on the vertical cross-range ($y$) and range directions.  
Finally, the third CV-Conv layer has a simple kernel size of $3 \times 3 \times 3$. 
For this simple example CNN, larger kernel sizes demonstrated improved numerical performance on the training and testing data. 
However, many CV-CNN architectures have been successfully employed with smaller kernel sizes in previous work \cite{jing2022enhanced,gao2018enhanced,dai2021imaging,fan2021fast}. 
The model is trained using an ADMM optimizer for 25 epochs with a learning rate of $1 \times 10^{-3}$, $\beta_1 = 0.9$, and $\beta_2 = 0.999$.

To demonstrate the ability of the trained network to generalize for real data, a planar scan is performed of a kitchen knife, shown in Fig. \ref{fig:knife_ex_cnn}b, using a mmWave radar. 
The image in Fig. \ref{fig:knife_ex_cnn}a is reconstructed using the planar RMA detailed in Section \ref{subsubsec:rma} before being processed by the CV-CNN. 
Even with a simple network, a considerable visual improvement can be noted in the super-resolved (SR) image in Fig. \ref{fig:knife_ex_cnn}c. 
The spatial features of the knife are retained. 
In particular, knife blade is significantly better resolved in the $z$-direction and the circular metal insets of the knife handle are much less obscured by distortion and noise compared with the input image. 
The ability of the network to generalize to real data demonstrates the feasibility of training data-driven image processing algorithms using synthetic datasets created by the proposed THz imaging toolbox. 
Generalization of training data synthesized by our toolbox is investigated across several frequency bands in \cite{smith2023dual_radar}.  
The flexibility of our simulation software allows for rapid generation of meaningful datasets for a variety of imaging problems beyond super-resolution, including object detection \cite{su2023object}, image segmentation \cite{shen2008detection}, scene understanding, and keypoint detection. 
Improving accessibility of THz training datasets will increase the acceleration of SAR image processing research and adoption in industry as the burden of simulating realistic scenes is eased by our simulation platform. 
Earlier versions of the proposed toolbox and interactive user interface have already gained traction for algorithm development \cite{smith2022efficient}, dataset generation \cite{smith2021An,smith2022ffh_vit,vasileiou2022efficient,smith2023dual_radar}, and prototyping \cite{gezimati2022circular,gezimati2023curved,smith2020nearfieldisar}, and are being made publicly available as an open-source software to foster a broader understanding of systems and algorithms for near-field mmWave and THz SAR and improve accessibility to essential tools for developing and benchmarking data-driven algorithms. 

\section{Conclusion}
\label{sec:conclusion}
With the emergence of 5G, 6G, and IoT technologies, mmWave and THz sensor suites have rapidly evolved and are trending toward ubiquity. 
Enabled by advancements in high-frequency hardware, the frequency ranges spanned by mmWave and THz devices are suitable for a wide range of applications in sensing and imaging. 
Synthetic aperture techniques leverage the large bandwidths achieved by such systems to enable ultra-high-resolution spatial sensing. 
Imaging with sub-millimeter resolutions in the mmWave and THz frequencies enables safer and more robust material characterization and NDT. 

In this article, we presented a tutorial review of systems and algorithms for THz SAR in the near-field with an emphasis on emerging algorithms that combine signal processing and machine learning techniques. 
We included an overview of classical and data-driven THz SAR algorithms, focusing on object detection for security applications and SAR image super-resolution. 
Relevant issues, challenges, and future research directions for emerging algorithms and THz SAR were detailed, including standardization for system and algorithm benchmarking, inclusion of state-of-the-art deep learning techniques, signal processing-optimized machine learning, and hybrid data-driven signal processing algorithms. 
Further investigation into these directions will facilitate the transformation of THz SAR from theory to practical application and improve numerous sensing modalities. 
Standardization and benchmarking are crucial issues that must be addressed for mmWave and THz SAR algorithms to flourish in the modern data-driven era. 
Finally, we introduced an interactive user interface and accompanying software toolbox for the exploration of THz SAR concepts, system prototyping, accelerated algorithm development, and efficient dataset generation. 
The proposed toolbox is employed to generate a realistic dataset, which is used to train a CV-CNN for image super-resolution. 
The resulting model is robust on real data, demonstrating the ability of the proposed toolbox to create meaningful, realistic datasets for mmWave and THz image processing applications. 
The tutorial review and research outlook on SAR imaging algorithms presented in this article  serve as a helpful reference and practical guidance for further in-depth exploration of emerging SAR algorithms and development of novel SAR systems. 

\appendices
\section{Downloading the Toolbox and Accessing the Documentation}
\label{app:documentation}
The proposed toolbox and interactive GUI can be downloaded via MathWorks or the authors' website, as listed below. 
Using the MathWorks file exchange platform, our toolbox can be downloaded and installed in MATLAB. 
To use the functionality of the toolbox from the repository available on our website, add the the main folder to the MATLAB search path.

\begin{itemize}
    \item \href{https://www.mathworks.com/matlabcentral/fileexchange/}{www.mathworks.com/matlabcentral/fileexchange/[to be released]}
    \item \href{https://github.com/josiahwsmith10/THz-and-Sub-THz-Imaging-Toolbox}{labs.utdallas.edu/wislab/projects}
\end{itemize}

To set up and open the directory of the toolbox, call the provided script \texttt{THzSimulator()}. 
(Note: The main folder of the toolbox must be on the MATLAB path and is included when installed via MathWorks). 
To open the interactive GUI, either call \texttt{THzSimulator()} and open the application in MATLAB Application Designer or call the \texttt{THzSimulatorGUI()} script to open the stand-alone GUI. 

To access the documentation created for this toolbox, after installing the packaged MATLAB toolbox or downloading the repository and adding the main folder to the MATLAB path, navigate to the \textit{Home} tab in the main MATLAB window and select the \textit{Help} button indicated by a question mark icon. 
The \textit{Documentation Home} window will open, and the THz Imaging Toolbox will appear in the \textit{Supplemental Software} section\footnote{Refer to \url{www.mathworks.com/help/matlab/matlab_prog/display-custom-documentation.html} for more detailed instructions.}. 
Additionally, the \textit{Info} menu and \textit{About} tab of the interactive GUI contain quick links to the documentation.
Inside our documentation are the Getting Started Guide, User Guide, Reference, and Examples sections detailing the usage and functionality of the various tools.

\bibliography{mega_bib}
\bibliographystyle{IEEEtran}

\end{document}